\newif\ifsiam 
\siamtrue
\siamfalse 
\ifsiam
\documentclass[review,onefignum,onetabnum]{siamonline171218}
\usepackage{float}
\else
\documentclass[leqno]{article}
\usepackage{graphicx}
\DeclareGraphicsExtensions{.eps}
\usepackage[letterpaper,left=3cm,top=3.5cm,right=3.5cm,bottom=6cm]{geometry}
\usepackage[pdftex,pdfpagelabels,bookmarks,hyperindex,hyperfigures,colorlinks=true,linkcolor=blue,citecolor=green]{hyperref}
\usepackage{amsmath}
\usepackage{amssymb}
\usepackage{pdfpages}
\usepackage{amsthm}
\usepackage{float}

\usepackage[utf8x]{inputenc}

\fi
\ifsiam

\input{shared_pap}
\ifpdf
\hypersetup{
  pdftitle={Dissecting the snake: transition from localized patterns to spike solutions},
  pdfauthor={N. Verschueren and A. Champneys}
}
\fi

\fi

\DeclareMathOperator\arctanh{arctanh}

\newcommand{\eps}{\varepsilon}
\newcommand{\beq}{\begin{equation}}
\newcommand{\eeq}{\end{equation}}
\begin{document}

\ifsiam

\maketitle

\else
\begin{center}

  {\Large
    Dissecting the snake: the transition from localized patterns to isolated spikes in pattern formation systems}
\vspace{0.3cm}

{\small Nicolas Verschueren\footnote{Department of Engineering
    Mathematics, University of Bristol, Bristol BS8 1TR, United Kingdom
    (n.verschuerenvanrees@bristol.ac.uk)} and Alan Champneys\footnote{
    Department of Engineering Mathematics, University of Bristol, Bristol BS8 1TR, United Kingdom
    (A.R.Champneys@bristol.ac.uk)}.}
\end{center}
\hrule
\fi

\ifsiam
\begin{abstract}
  \else
\begin{center}
\textbf{Abstract}\\
\fi

An investigation is undertaken of coupled reaction-diffusion systems
in one spatial dimension that are able to support, in different
regions of their parameter space, either an isolated spike solution,
or stable localized patterns with an arbitrary number of peaks.  The
distinction between the two cases is characterized through the behavior of
the far field, where there is either an oscillatory or a monotonic
decay. This transition is illustrated with two examples: a generalized
Schakenberg system that arises in cellular-level morphogensis and a
continuum model of urban crime spread.  In each, it is found that
localized patterns connected via a so-called homoclinic snaking curve
in parameter space transition into a single spike solution as a second
parameter is varied, via a change in topology of the snake into a
series of disconnected branches. The transition is
caused by a so-called Belyakov-Devaney transition between complex and
real spatial eigenvalues of the fair field of the primary pulse. A
codimension-two problem is studied in detail where a non-transverse
homoclinic orbit undergoes this transition. A Shilnikov-style analysis
is undertaken which reveals the asymptotics of how the infinite family
of folds of multi-pulse orbits are all destroyed at the same parameter
value.  The results are shown to be consistent with numerical
experiments on the examples.  \ifsiam
\end{abstract}
\else
\end{center}
\fi

\ifsiam
\begin{keywords}
  \else
  \textbf{Key words:}
  \fi
 Reaction-diffusion, localized patterns, homoclinic snaking,
 Belyakov-Devaney, Shilnikov analysis
 \ifsiam
\end{keywords}
\else
\\
\fi

\ifsiam
\begin{AMS}
  \else
  \begin{flushleft}
  \textbf{AMS subject classifications.}
  \fi
  35B25, 35B32, 35K57, 34B07\\
  \ifsiam
\end{AMS}
\else
\end{flushleft}
\fi
\section{Introduction}
\emph{Localized structures} are a common feature exhibited by
spatially extended systems far from equilibrium. By a localized structure
we mean a non-trivial pattern that is localized to some portion of
the domain, with exponential decay in the far field.  Such structures
can be found in diverse contexts at different spatial scales ranging
from biochemistry to planetary physics. Excellent reviews on localized
structures can be found in \cite{purwins,revoptls,Knoblochreview}.

Broadly speaking, it is possible to distinguish two types of spatially
localized structures: those without and those with a distinguished
spatial wavelength. The two cases are sometimes referred to
respectively as isolated {\em spikes} (or pulses) and {\em localized
  patterns} (patches of the domain within which there is a spatially
modulated periodic structure) respectively.

In this paper, we restrict our attention to the simplest setting in
which one might see such localized structures: systems of two
reaction-diffusion equations in one spatial dimension. Specifically,
we study partial differential equation (PDE) systems of the form
\begin{equation}
\partial_t U= D U_{xx} +F(U,U_x;\lambda),
\label{eq:boringpde}
\end{equation}
where $x\in [-L,L]$ with $L\gg 1$ and where the state of the system
is represented by the vector unknowns 
$U \in \mathbb{R}^2$ and $\lambda\in \mathbb{R}^p$ is a vector of 
parameters. Here $D$ is a diffusion matrix and we assume that the 
local kinetic function $F$ is odd in components of $U_x$, so that the 
system is invariant under $x \to -x$. 

\begin{figure}
\begin{center}
\includegraphics[width=\textwidth]{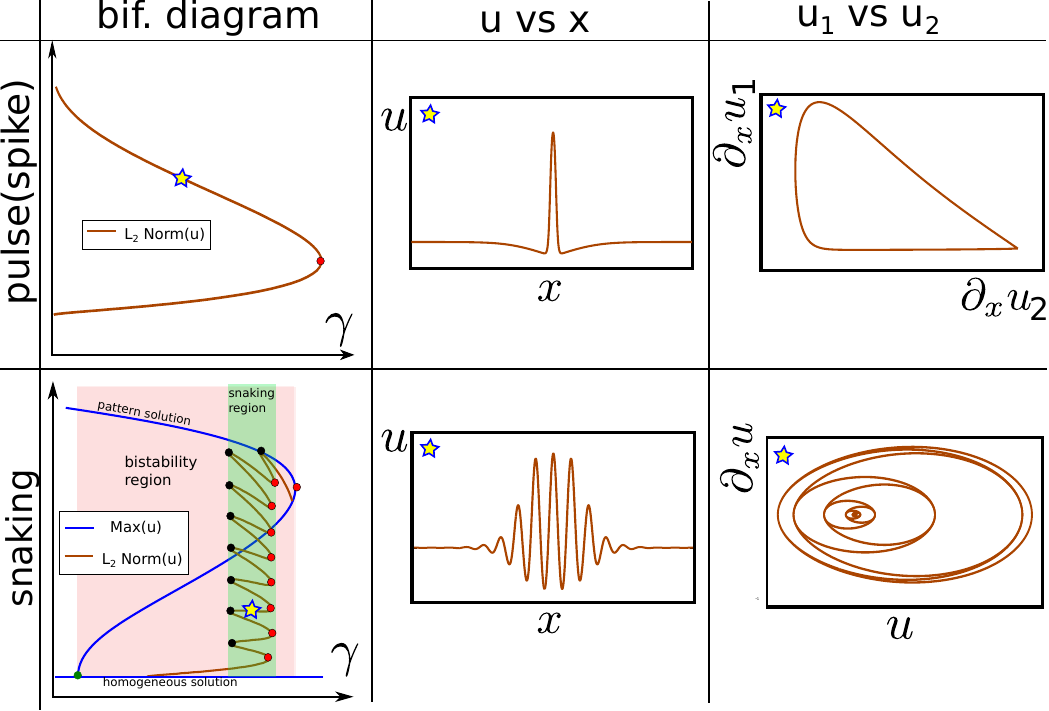}
\end{center}
\caption{Qualitative differences between pulse solutions (top half) and localized patterns (bottom
  half). In the \emph{leftmost} column, a sketch of the bifurcation diagram
  of each solution is represented in which the vector $L_2$-norm is
  plot against a scalar parameter $\gamma$. The
  \emph{middle} column depicts one representative component of a
  particular solution (distinguished with a star on the left), as a
  graph against $x$. The \emph{rightmost} column shows 
a planar projection of the solution in phase space. [Numerical
results used here are from: (pulses(spike)) the generalized cell polarity model (\ref{dimle}) below with parameter values (\ref{eq:pinpars}) using $\delta$ as a continuation
parameter when $\gamma=14$; (sanking)  using the Swift-Hohenberg equataion $\partial_t
u=-(1+\partial_x^2)^2 u-\mu u +\nu u^2-u^3$ on a finite domain with $\nu=1.6$ and $\mu$ as 
the continuation parameter.]}

\label{fig:schess}
\end{figure}

Where convenient, we shall consider the limit $L \to \infty$ and study the
the steady problem of (\ref{eq:boringpde}) in the
context of {\em spatial dynamics}, in which the spatial variable $x$ is
considered to be a time-like co-ordinate (see
e.g.~\cite{ioosbook}). In this context, the steady problem to
\eqref{eq:boringpde} may be considered to be a four-dimensional
reversible system in the phase space variables $(U,U_x)$ and
a localized structure corresponds to an orbit $U_h(x)$ that is 
homoclinic to the homogeneous equilibrium solution $U_0$ as
$x\to \pm \infty$. One can then use the theory of homoclinic orbits
in reversible system, see e.g. \cite{Devaney2,alansnake,apd,snakebeck}, which
shows that such localized solutions are persistent under parameter variation.
Very different kinds of behavior can be observed however, depending on
the nature of the spatial eigenvalues of the problem linearized about
$U_0$. Broadly speaking, real eigenvalues correspond to isolated
spike-like solutions whereas complex eigenvalues lead to the possibility of
localized patterns. See Figure~\ref{fig:schess} which illustrates the 
qualitative differences between these two cases. 

\subsection{Isolated spikes}

Spike-like isolated or single-pulse states correspond to
\emph{homoclinic} steady solutions of the system
(\ref{eq:boringpde}) in which the decay to the far field is eventually
monotonic and their representation in the spatial phase space is
rather simple (see the upper panels of Figure.~\ref{fig:schess}).
Stable versions of such pulses often arise via a single fold
bifurcation from an unstable pulse that bifurcates at small amplitude
from a homogeneous equilibrium (see the bifurcation diagram sketched
in the \textsl{top-left} panel of Figure \ref{fig:schess}).

Such solutions can be observed in a wide variety of models.
Two important examples are Gray-Scott or Schnackenberg-like 
reaction-diffusion systems ---
which can be analyzed using either geometric singular perturbation
theory (as, for example, in the works of Doelman et
al.~\cite{doeloriginal,fritsexplicit}) or using matched asymptotic
expansions (as in the work of Ward {\em et al.}
\cite{wardfirst,wardrecent}) --- and the parametrically driven
non-linear Schr\"odinger equation (see, for example, the works of
Barashenkov {\em et al.} \cite{barafamous,brara2} and references
therein). In some contexts such localized states are called \emph{dissipative
  solitons} (see e.g.~\cite{Akhmediev,purwins}) and methods exist to
approximate their dynamics on large domains via point-like
approximations (see e.g. \cite{Promislow}).  In two or more spatial
dimensions such solutions can behave like particles and can weakly
interact with each other. Although some mechanisms exist (such as
so-called inclination or orbit-flips, see
e.g.~\cite{Sandstede,Oldeman} and references therein) 
for creating stable multi-peaked versions,
in general such localized states do not form infinite
families of bound states, owing to their lack of oscillatory
tails. Also, unlike solitons of integrable systems, dissipative
solutions can gain permanent oscillatory dynamics,
for instance through Hopf bifurcations, which
can give rise to quasi-periodic or fully chaotic behaviors.

\begin{figure}
\begin{center}
\includegraphics[width=\textwidth]{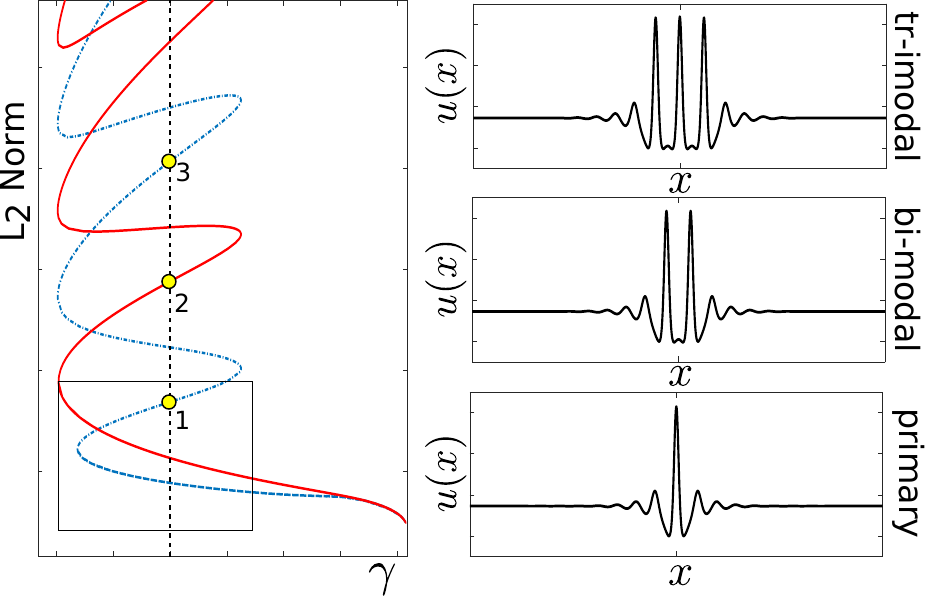}
\end{center}
\caption{More details of the homoclinic snake in Figure.~\ref{fig:schess}, 
showing two inter-twined branches; the  
continuous and dashed lines represent branches of 
symmetric solutions with even and odd number of large peaks,
respectively (without considering the oscillating tails). The square in
  the bottom of left panel, is the area where the primary (single peaked)
  orbit exists. One component of the solution profile at each numerical label is depicted in the corresponding graph to the right.
For both curves in the main graph, the localized solutions are stable
along portions of the graph that have positive slope and lose stability
at the fold points. }
\label{fig:double_snake}
\end{figure}

\subsection{Localized patterns and homoclinic snaking}
\label{sec:local-patt-snak}
A localized pattern represents a qualitatively different 
type of
stationary solution state to \eqref{eq:boringpde}. In contrast to spikes,
such states possess a spatial wave-length and feature decaying
oscillatory tails as the solution tends to the far field.
These solutions typically occur as part of an
infinite family of solutions, members of which can be characterized by
their number of maxima outside of the tail regions (see the lower panels in
Figure~\ref{fig:schess} and also Figure~\ref{fig:double_snake}).

One way to describe such localized patterns is to think of them as the
interaction of two fronts, where at least one of the steady
states that the fronts are connecting posses a spatial
wave-length \cite{Claudiohole}.  In this situation, it is sometimes
possible to use a weakly non-linear analysis to estimate the
bifurcation diagram of localized patterns (e.g. \cite{miophil}), which
typically resembles the one illustrated in the \textsl{bottom-left}
panel of Figure \ref{fig:schess}.  This technique assumes a large
distance between the fronts.

An alternative way of understanding localized patterns via spatial
dynamics assumes instead that the localized patterns are not weakly
interacting, but arise from the unfolding a heteroclinic connection
between a saddle-focus equilibrium and a saddle-like periodic
orbit. The period of the periodic orbit provides the internal
wavelength of the patters.  As a consequence of the reversibility, the
unfolding of the heteroclinic connections creates a countable infinite
number of homoclinic orbits, which correspond to the localized
patterns. The simplest examples among these infinite families are
typically connected in a bifurcation diagram like the one depicted in
Figure.~\ref{fig:double_snake} which has been dubbed a {\em homoclinic
  snake} \cite{alansnake}.

There is now a rich literature on the homoclinic snaking scenario. For
example, as explained in the context of Swift-Hohenberg-like equations
in the work of Burke \& Knobloch \cite{burke01,burke02}. Kozyreff \&
Chapman \cite{Kozyreff,Kozyreff2} among others (e.g.~\cite{Dean}),
have explained how the snake arises from beyond-all-orders
perturbations of a degenerate (super-to-subcritical transitioning)
pattern formation instability of the homogeneous steady state.  In
addition, Beck {\em et al} \cite{snakebeck} have provided rigorous
justification of what can happen away from this singular limit
(see also \cite{Makrides} for more cases). In particular, it is
important to draw a distinction between cases where there is and where
there is not variational structure (a conserved Hamiltonian-like
quantity of the spatial dynamics), whether or not there is additional
symmetry and also between finite and infinite domains;
e.g. \cite{Dawes2,BurkeDawes,Houghton1,Houghton2,Dawes,norevsnake}.

There are also analogues of homoclinic snaking in higher spatial
dimensions, where localized versions of rolls, hexagonal lattices and
target patterns can be observed \cite{2Dsnake,Lloyd2}. %
The snaking diagrams are more complex, and not all details are
known, but see \cite{McCalla,Kozyreff3,Bordeu} for the state of the
art.  For the preset paper, though, we shall exclusively concern
models of the form \eqref{eq:boringpde} in one spatial dimension.

\subsection{Outline}
This paper has been motivated by a number of recent studies in which
both localized patterns and isolated spikes have been observed in
different parts of parameter space of reaction-diffusion systems of
the form \eqref{eq:boringpde}. One motivation for our study is the
work of Zelnik {\em et al} \cite{yuvalss} on an ecological model, in
which exactly the same transition we study in this paper is observed,
although a theoretical explanation of the spike to localized pattern
transition was lacking.  Similar transitions have been observed in our
recent work on a simple model for cellular polarity formation
\cite{miosiads} and in a continuum model for urban crime \cite{crimelloyd}. These two models will be revisited in detail in
Section~\ref{sec:examplesss}.

The rest of the paper is organized as follows. Through the two
examples discussed in section \ref{sec:examplesss}, a hypothesis is
formed that the transition from localized patterns to isolated spikes
is driven by where the first fold of the homoclinic snake passes
through the curve in a parameter plane where there is a transition
from complex to real spatial eigenvalues, through a double real
eigenvalue. We dub such a codimension-two bifurcation a non-transverse
Belyakov-Devaney bifurcation.  Section \ref{sec:shiln-type-analys} 
contains the main results of the paper, a partial unfolding of such a
non-transverse Belyakov-Devaney bifurcation, using a Shilnikov-style
approach --- an adaptation to the analysis in \cite{subsidiaryalan} to
the particular codimension-two problem in question. Section \ref{sec:asymptotics} than makes asymptotic predictions from
the analysis on the nature of the bifurcation and compares
these to results for the two examples.  The paper ends with a careful
discussion in Section~\ref{sec:discussion}, indicating exactly what has and
has not been shown, and pointing to directions of future work.

\section{Snake to spike transition in two examples}\label{sec:examplesss}

We have chosen two example systems of the form \eqref{eq:boringpde} to
illustrate the phenomenon we seek to explain.  As we shall see, these
models have very different nonlinear terms and arise in quite distinct
contexts, which serves to illustrate the ubiquity of
the transition we seek to understand (see also
Section~\ref{sec:discussion} for discussion of further examples).

The steady problem of (\ref{eq:boringpde}) can be re write as a four-dimensional, reversible
system of ordinary differential equations
\begin{equation}
  \label{eq:spaedo}
\frac{d y}{dx} =f(x,\lambda), \; \;y \in \mathbb{R}^4 = (u,u_x,v,v_x).
\end{equation}
As first shown by Devaney \cite{Devaney1,Devaney2} (see also the
reviews \cite{apd,apd2}), the multiplicity of homoclinic orbits to a
symmetric equilibrium $y_0$ in such a systems depends crucially on the
eigenvalues of the linearization around $y_0$. Reversibility ensures
eigenvalues come in symmetric pairs $\pm \lambda$. If the equilibrium
is hyperbolic (all eigenvalues have non-zero real parts) then a
generic homoclinic orbit will be isolated and should persist under
parameter perturbation.  If the eigenvalues are complex $\lambda =\pm
\rho \pm i \omega$, then each homoclinic orbit should be accompanied
by an infinite multiplicity of families of $N$-pulse homoclinic
orbits. For each $N$, the family is characterized by a string a $N-1$
integers $(i_1,,i_2,\ldots i_{N-1})$, where each $i_k \in
\mathbb{Z}^+$ represents the distance between each pulse in terms via
the number of half-oscillations close to $y_0$.  There can be
restrictions on which strings are admissible, that is which ones
correspond to true multi-pulse orbits. For example, without a
conserved first integral then only palindromic strings, which
correspond to symmetric orbits, will lead to persistent multi-pulse
orbits; see \cite{Harterich,subsidiaryalan} for details.

\begin{figure}
\begin{center}
\includegraphics[width=\textwidth]{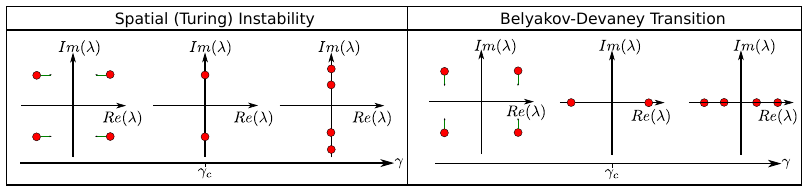}
\end{center}
\caption{Eigenvalues of the linearization around a homogeneous
  solution of system (\ref{eq:spaedo}) in the vicinity of a Turing instability (\emph{left panel}) and Belyakov-Devaney
  transition (\emph{right panel}). The arrow at the bottom shows the parameter $\gamma$ crossing a critical value $\gamma_c$.}
\label{fig:transitionsss}
\end{figure}

Homoclinic snaking, as in Figure.~\ref{fig:double_snake}, typically
occurs when a primary homoclinic orbit emerges sub-critically from a
point of double imaginary eigenvalues $\pm i \omega$. Such a
codimension-one bifurcation is sometimes called a Hamiltonian-Hopf
bifurcation, or reversible 1:1 resonance point, and is equivalent to
the fundamental pattern formation instability of the underlying
PDE, which, as the domain size tends to
infinity, corresponds to the accumulation point of infinitely Turing
bifurcations \cite{Victor2}.  With slight abuse of notation, we shall
refer to such a double-imaginary eigenvalue bifurcations in the
spatial dynamics (cf. left panel of Figure \ref{fig:transitionsss}) as
a {\em Turing bifurcation}. Note that the localized patterned states
that arise from the fold bifurcations within the homoclinic snake are
typically quite distinct from the multi-pulse homoclinic orbits,
because their large peaks are not close to the primary homoclinic
orbit, but to a finite amplitude periodic orbit; and the separations
between the peaks is governed by the period of the periodic orbit, not
by the linearization at the equilibrium.  Nevertheless each localized
pattern arising in the snake will give rise to infinitely many
further multi-pulse orbits just like the primary orbit does.

The central theme of this paper is how homoclinic
snaking can break up, as a second parameter is modified; specifically what happens 
when the pinning region extends to cover the whole of a parameter region in
which the spatial eigenvalues are complex. The boundary case of interest
is when these eigenvalues become a double real pair, see the right panel of
Figure~\ref{fig:transitionsss}. The case of primary homoclinic orbit
passing through such a transition was dubbed in \cite{apd} a {\em
  Belyakov-Devaney} bifurcation because of the similarity to the
codimension-two bifurcation in non-reversible systems first described
by Belyakov \cite{Belyakov1}; see also \cite{Belyakov2} for these
ideas applied in the context of reversible systems.  Here, the
infinite family of multi-pulse orbits disappear via the distance
between each pulse tending to infinity as the critical eigenvalue
transition is approached. Where convenient, in what follows we shall
sometimes abuse notation and describe the eigenvalue transition of
four complex eigenvalues to four real eigenvalues through a
double-real-eigenvalue transition as being a {\em Belyakov-Devaney
  transition}, irrespective of whether the equilibrium has a
homoclinic orbit connecting to it or not.
What we shall find though is that as the folds of the homoclinic snake
approach a Belyakov-Devaney transition, then the distinction between a
multipulse homoclinic orbit and a localized pattern becomes blurred as
the period of the periodic orbit at the center of the pattern tends to
infinity and the pattern itself becomes part of the family of
multi-pulse orbits associated with the primary homoclinic orbit.

Notice that is often possible to find analytical expressions for the
Turing and Belyakov-Devaney transitions, because they are properties
of the linearized system (see examples below). However, in order to
study the existence and bifurcations of homoclinic orbits away from special
distinguished points, the use of numerical methods is required. Throughout
this paper, we shall use AUTO \cite{auto} with Neumann boundary
conditions on a half-interval $[-L,0]$ or $[0,L]$ for a sufficiently
large $L$ that the solution in the far field is very close to the
homogeneous equilibrium. Using this approach, we will only capture
homoclinic orbits that are symmetric under the reversibility.
Also, this paper shall only consider existence and multiplicity of homoclinic
solutions, the stability and PDE dynamics associated with such localized
solutions shall not be of specific concern. 

\begin{figure}
\begin{center}
  \includegraphics[width=0.9\textwidth]{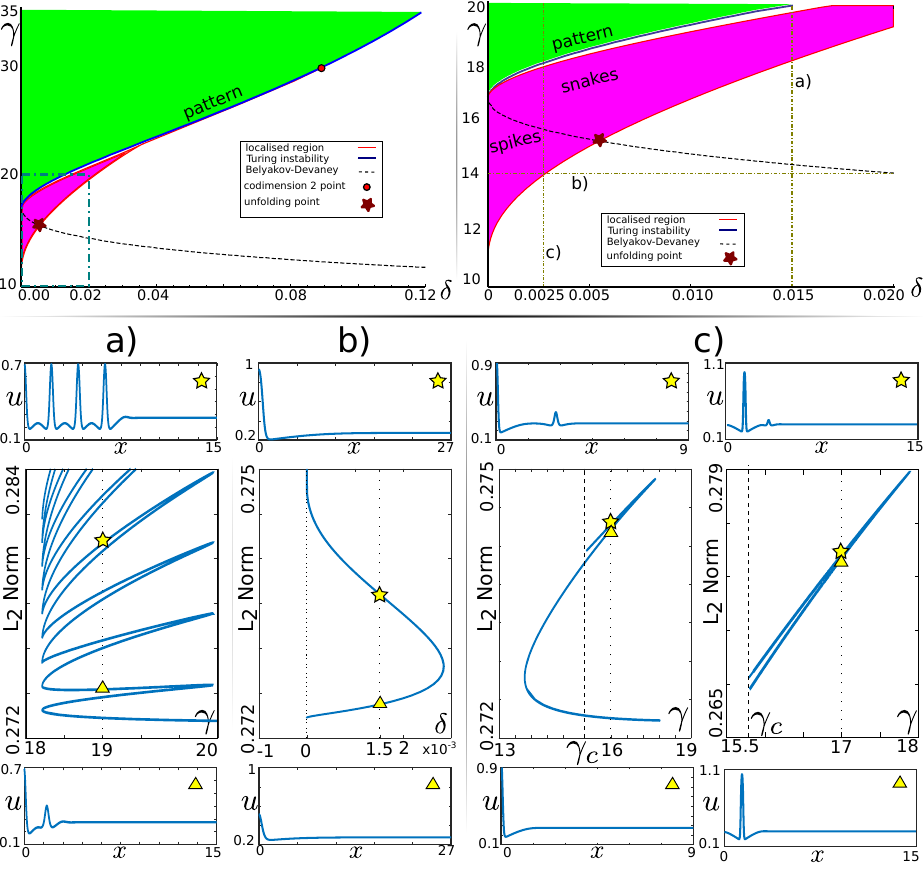}
\end{center}
\caption{\textsl{Top panel}: Two-parameter bifurcation
  diagram of the general cell polarity model
  \eqref{dimle} with other parameters as in \eqref{eq:pinpars}. 
 The right-hand panel is a zoom of the area within the
  dot-dashed box in the left-hand panel. The solid (blue) line
  corresponds to the Turing bifurcation which is sub-critical to the
  left of the red circle and super-critical to the right. The dashed
  line corresponds to the Belyakov-Devaney transition. The thin red lines 
delineating the pink shaded region correspond to where localized patterns or
spikes are observed. The codimension-two point of concern in this paper
is represented by the maroon star at the 
intersection of the Belyakov-Devaney line and the right-hand edge of
the localized pattern region.
Alphabetically labeled dot-dashed lines in the 
right-hand panel correspond to one-parameter continuations depicted in
the \textsl{Bottom panels}, insets to which show solution profiles depicted
on the half-interval $x\in [0,L]$ for $L\gg 1$; see text for details.}
\label{fig:pmap_wp}
\end{figure}

\subsection{A Generalized Cell Polarity Model}
\label{sec:cell}
The first example is the model, arising in cell biology,
that was previously studied by the present
authors in \cite{miosiads}. This is a generalization, through the addition of
source and loss terms, of the simple
canonical model for the spatial patterning of G-proteins underlying
cellular polarity formation proposed by Mori, Jilkine \& Edelstein-Keshet
\cite{Morior}.  It can be written in the form
\begin{subequations}
\label{dimle}
\begin{align}
\label{dimlea}
\partial_t u &=\delta\partial_{xx}u +[F(u,v)-\varepsilon \theta u],\\
\label{dimleb}
\partial_t v &=\partial_{xx}v - [F(u,v)-\varepsilon \alpha],\qquad x\in 
\left[ -L,L \right], \quad \partial_x (u,v) 
(\pm L)=0 , \\
\label{dimlec}
\text{where} & \quad F(u,v) =\gamma \frac{u^2v}{1+u^2}-\eta u+v. \qquad
\end{align}
\end{subequations}
Here $u(x,t)$ and $v(x,t)$ represent the concentrations of active and
inactive species respectively of a structural $G$-protein, and
$\delta \ll 1$ is the ratio of their diffusion rates.  The function $F(u,v)$
represents the local kinetics of the activation step parameterized by
$\mathcal{O}(1)$ parameters $\eta$, $\gamma$.  The specific form of
$F$ is not important, provided that it exhibits bistability.  
We fix all the parameters except two, the
diffusion ratio $\delta$ and the parameter which controls the
non-linearity in the model $\gamma$. Following \cite{miosiads}, 
we choose
\begin{equation}
\varepsilon=1, \quad \eta=5.2, \quad \theta=5.5, \quad \alpha=1.5.
\label{eq:pinpars}
\end{equation}
The unique homogeneous equilibrium of \eqref{dimle} is given by
\begin{equation}
(u_0,v_0)=\left(\frac{\alpha}{\theta},\frac{\alpha (\varepsilon \theta +\eta)(\theta^2+\alpha^2)}{\theta[\theta^2+\alpha^2(1+\gamma)]}\right):=\left(\frac{\alpha}{\theta}, \beta_0+\varepsilon \beta_1\right).
  \label{homoeq}
\end{equation}
Performing a linear stability analysis around this equilibrium, the
condition for the linear transitions in question are \cite{miosiads}:
\begin{equation}
  \label{eq:critwp}
\varepsilon \theta \partial_v F -\frac{ (\partial_u F -\delta \partial_v F -\varepsilon \theta)^2 }{4\delta}  =0, \qquad \mbox{with }
k_c^2=\frac{\partial_u F-\delta \partial_v F-\varepsilon
  \theta}{2\delta} ,
\end{equation}
with critical wavenumber
$k_c^2>0$ corresponding to the
Turing instability and $k_c^2<0$ to 
a Belyakov-Devaney transition. The double pairs of spatial eigenvalues
at the Turing
bifurcation are $\pm i \omega$ where $\omega=k_c$, and
$\pm \lambda=\mp k_c$ at the Belyakov-Devaney transition.  

The organization of simplest patterned states of model
(\ref{dimle}) in the $(\delta,\gamma)$-plane
is summarized in Figure~\ref{fig:pmap_wp} (full details are given in
\cite{miosiads}). The top panel depicts the single curve
(\ref{eq:critwp}) for the case $k_c^2>0$ and $k_c^2<0$ 
using a continuous blue line and a
dashed black line respectively. The red dot represents the point
where the Turing bifurcation changes from supercitical
(for larger $\delta$) to subcritical. From this point emerges
a pinning region, shaded pink, in which localized solutions occur.
A maroon star marks the
codimension two point where the Belyakov-Devaney transition and the
first fold of the primary homoclinic orbit (right-hand boundary
of the localized pattern region) occur simultaneously.
This point, which we term the \emph{unfolding
  point}, will be the subject of our study in Section
\ref{sec:shiln-type-analys}.

The lower panels of Figure~\ref{fig:pmap_wp}
depict various cross-sections through the bifurcation diagram of
the simplest, {\em primary} homoclinic orbit, 
with a yellow triangle and star used to depict two specific homoclinic solutions plot on the
half-interval $x >0$
Panel (a) shows a regular
homoclinic snake, for a case where $\gamma$ remains entirely in the
parameter region corresponding to complex eigenvalues. There are
infinitely many folds in theory but the simulation is always carried
out on a finite domain.  

In contrast, panel (b)
shows a case where localized patterns are contained 
exclusively in the real eigenvalue part of parameter space.
In \cite{miosiads}, the only stable
localized solution observed in this parameter region is the
larger-amplitude single spike solution observed on the upper
portion of this branch. 
Note that both the small and
large amplitude pulse survive all the way to the singular limit
$\delta \to 0$, with the latter disappearing at zero amplitude
deviation from the homogeneous equilibrium.

Panel (c) of Figure \ref{fig:pmap_wp}
shows a hybrid case where the primary localized solution
crosses the Belyakov-Devaney transition, at $\gamma=\gamma_c$,
indicated by a vertical dashed line. The figure is divided into two separate
portions for the primary orbit and the two-pulse orbits that were formerly
part of the homoclinic snake. Note how 
the first fold of the primary orbit has passed through
$\gamma_c$. After the second fold, which creates a three-peaked orbit,
the primary branch turns around and then terminates at $\gamma_c$. The
mechanism of termination is that the solution becomes delocalized; the
outer two pulses disappear towards $x=\pm \infty$ as $\gamma \to
\gamma_c$. We find that all other branches of the snake, including the
intertwined branch with an even number of pulses never
cross $\gamma_c$, but are split into separate pieces with solutions
becoming delocalized as they approach $\gamma_c$. The second
portion of panel (c) shows the continuation of the two-pulse
orbit which exists only for
$\gamma>\gamma_c$. Note how both the primary and the two-pulse
homoclinic orbit gain an extra small maximum after returning from
their right-most fold, which also becomes delocalized as  
$\gamma \to \gamma_c$.

\subsection{An urban Crime-Wave model}
In 2008 Short {\em et al} \cite{shortcrime} derived a system of 
PDEs for crime density within an urban area, 
based on mean-field
approximation to a stochastic agent-based model. The model was
able to replicate the behaviors observed in real crime data, in particular
that crime seems to be concentrated in localized areas
(hot-spots).  Subsequently, 
Lloyd and O'Farrell \cite{crimelloyd} studied the same model in both one
and two spatial dimensions in an attempt to understand the origin of such
localization. Here we just focus on the one-dimensional model,
which can be written in the form
\begin{subequations}\label{eq:crime_wave}
\begin{align}
\label{eq:crime_wavea}
\partial_t A &=\eta^2\partial_{xx}A-A+\beta+\rho A,\\
\label{eq:crime_waveb}
\partial_t \rho &=\partial_{xx}\rho -2\frac{\rho}{A} \partial_{xx}A -2 \frac{\partial_x \rho  \partial_x A}{A} +2\frac{\rho (\partial_x A)^2}{A^2}-\rho A +\mu-\beta.
\end{align}
\end{subequations}
\begin{figure}
\begin{center}
\includegraphics[width=0.9\textwidth]{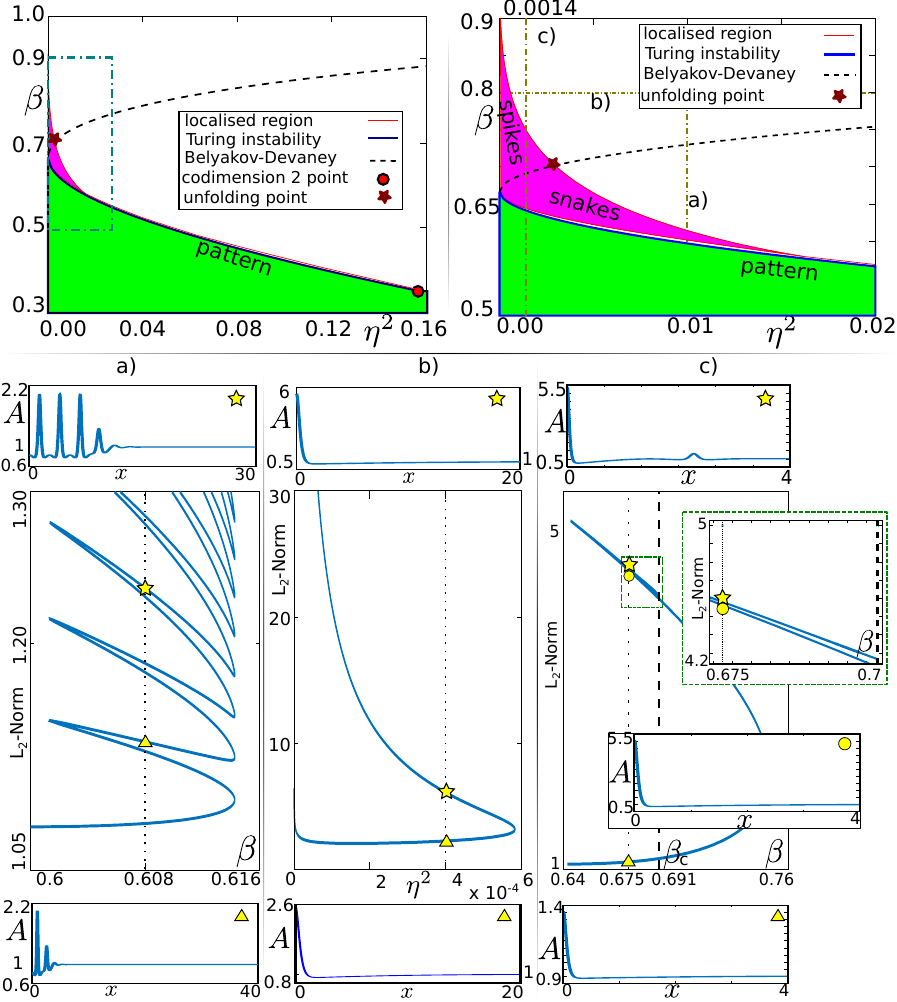}
\end{center}
\caption{\textsl{Top panel:} Two-parameter bifurcation diagram for the urban crime model 
  (\ref{eq:crime_wave}) for $\mu=1$, using the same conventions as in
  Figure.~\ref{fig:pmap_wp}. \textsl{Bottom panel:} One-parameter
bifurcation diagrams along paths a)--c)}
\label{fig:political_crime}
\end{figure}
Here $A$ represent the \emph{attractiveness} of an area to burglars and
$\rho$ the density of criminals. This system possesses three
dimensionless parameters, $\beta$, $\mu$ and $\eta^2$,
which, for the sake of clear distinction between parameters 
and state variables, 
we have renamed from the notation used in 
\cite{crimelloyd}. System \eqref{eq:crime_wave} has a homogeneous
equilibrium given by
$$
\left(\begin{array}{c}A_h\\
    \rho_h\end{array}\right)=\left(\begin{array}{c}\mu\\
      1-\frac{\beta}{\mu}\end{array}\right).
    $$
Performing a linear stability analysis around this equilibrium
\cite{crimelloyd}
one finds that a Turing instability occurs whenever 
\begin{equation}
\beta = \frac{2}{3}\mu-\eta^2\frac{\mu^2}{3}-\frac{2}{3}
\mu \sqrt{\mu\eta^2},\quad \text{with} \quad k_c^2 = \sqrt{\frac{\mu}{\eta^2}},
\end{equation}
and a Belyakov-Devaney transition when
\begin{equation}
\beta = \frac{2}{3}\mu-\eta^2\frac{\mu^2}{3}+\frac{2}{3}
\mu \sqrt{\mu\eta^2}.
\end{equation}

Figure \ref{fig:political_crime} presents a two-parameter bifurcation
diagram, using the same conventions and colors as the last example.
The bifurcation diagram was essentially already present
\cite{crimelloyd} (See Figure.~1(e) of that paper), but the transition
between spikes and localized patterns illustrated the bottom panel of
Figure \ref{fig:political_crime} was not analyzed.  Note how,
qualitatively speaking, the scenario observed here is identical to
that in the previous example.

\section{Unfolding the non-transverse Belyakov-Devaney transition}
\label{sec:shiln-type-analys}

The examples studied suggest that the transition we describe is
generic, to that end we now construct a plausible analysis, to predict
what happens in a neighborhood of the key codimension-two bifurcation
in question. That is, we provide a partial unfolding of the case of a
fold in a primary homoclinic orbit first touching
the Belyakov-Devaney transition and see if this can explain
the phenomenon observed.  We shall use the
paradigm of spatial dynamics, in the limit $L \to \infty$, in which we
consider the steady problem and the spatial variable $x$ is replaced
by a time-like variable $t$ in what follows (which should not be
confused with the actual temporal variable of the time-dependent
PDEs).

We use the method pioneered by L.P.~Shil'nikov, see \cite{shilnibook}
and references therein, specifically an adaptation to the previous
work of the second author \cite{subsidiaryalan}, that considered
symmetric homoclinic orbits to an equilibrium with complex eigenvalues
in a reversible system. Here, we will consider a perturbation of the
the primary homoclinic orbit, which will provide the conditions for
the existence of 2-pulse homoclinic orbits. The spirit of the
analysis is that of a formal, justified calculation; we shall not
attempt to provide rigorous statements.

\subsection{Generic hypotheses}
\label{sec:hypo}

We are interested in a description of homoclinic orbits in a
neighborhood of the codimension-two point where a primary homoclinic
orbit of a four-dimensional reversible system undergoes a fold at a
Belyakov-Devaney point. This point has been distinguished in the
two-parameter bifurcation diagrams of Figures \ref{fig:pmap_wp} and
\ref{fig:political_crime} with a maroon star and termed the
\emph{unfolding point}. In the spirit of \cite{subsidiaryalan}, we
consider the four-dimensional dynamical system given by the
differential equations
\begin{equation}
\label{eq:odesys}
\frac{d y}{dt}=f(y;\varepsilon,\nu)\quad \varepsilon, \nu\in
\mathbb{R},\quad y\in \mathbb{R}^4,\quad f\in C^{r}, \quad (r\geq 2).
\end{equation}
We shall make a number of generic hypotheses

\begin{description}
\item[H1]
We suppose that the system is reversible, that is, there is
a linear operator $\mathcal{R}$ which satisfies
$$ 
\mathcal{R} \circ \mathcal{R}= I \quad \text{and} \quad
\mathcal{R} \cdot f=-f\circ \mathcal{R}.
$$
It will be useful for our analysis to define the two-dimensional 
\emph{symmetric section} $S=\mbox{Fix}(\mathcal{R})$.
Whenever an orbit intersects $S$, then we can reverse time without
loss of generality so that the orbit
is symmetric, that is 
\begin{equation}
\mathcal{R} y(t)=y(-t) .
\label{eq:minus}
\end{equation}
\end{description}

\begin{figure}
\begin{center}
\includegraphics[width=\textwidth]{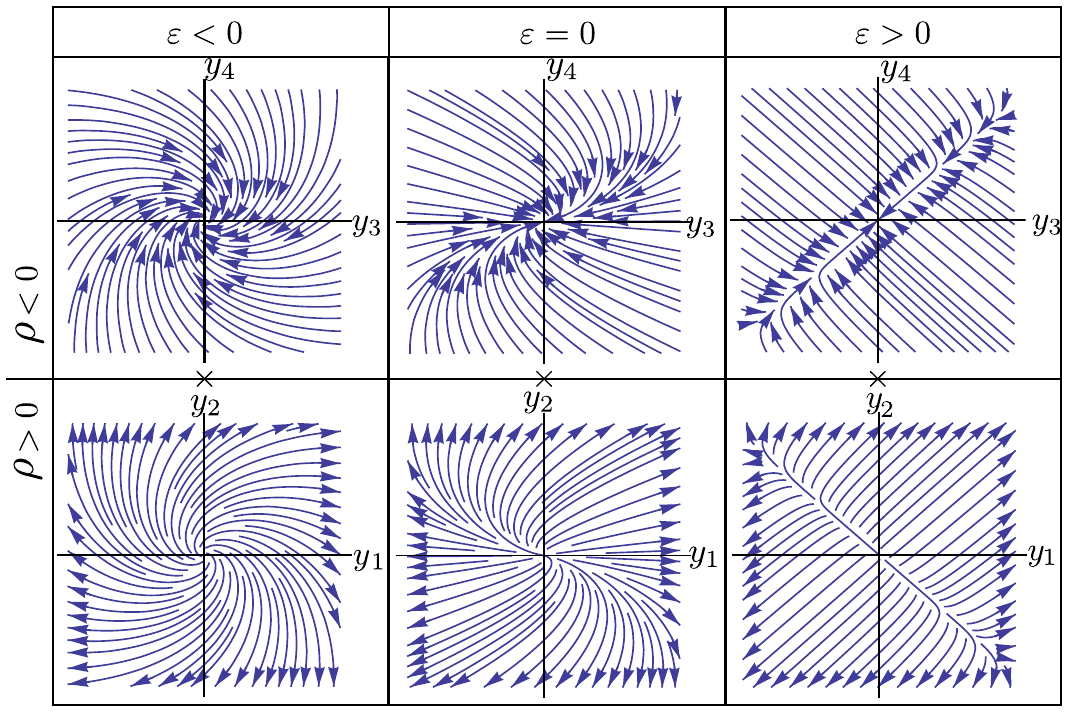}
\end{center}
\caption{Phase portraits of the linear dynamics within the stable and unstable
  eigenspaces a neighborhood $B$ of the origin, as described in using the local co-ordinates in Sec.~\ref{sec:locm}. In each column, the
  upper and lower panels correspond to stable and unstable manifolds,
  respectively. Chosen parameter values are $\rho=1$
     and
  from left to right $\eps=-0.8$, $0$ and $0.8$.}
\label{fig:pp}
\end{figure}

To apply the present analysis to  systems of two-reaction diffusion
equations, as in Section~\ref{sec:cell},
we have in mind that $y = (u,u_x,v,v_x)^T$ and that $\mathcal{R}$ acts to
reverse the derivative variables
$$
\mathcal{R} (u,u_x,v,v_x)^T = (u,-u_x,v,-v_x)^T \:,
$$
and similarly for the crime wave model. 
\begin{description}
\item[H2]
We suppose that the system (\ref{eq:odesys}) has an isolated hyperbolic
stationary point which lies within $\mathcal S$; for the sake of simplicity we take
this stationary point to be the origin $y=0 \in S$. We suppose that the linearisation at the stationary point is hyperbolic and has two-dimensional stable and unstable eigenspaces.  Then, by standard theorems on reversible systems (see e.g.~\cite{Devaney2,apd,Sevryuk}), the two-dimensional 
stable and unstable manifolds $W^{s,u}(0)$) 
$$
W^s(0)=\{y\in \mathbb{R}^4 | \lim_{t\to \infty}\phi_t(y)=0\}\quad\text{and}\quad W^u(0)=\{y\in \mathbb{R}^4 | \lim_{t\to -\infty}\phi_t(y)=0\}
$$
(where $\phi_t$ is the \emph{flow} corresponding to the differential equation)
are symmetric images of each other, $\mathcal{R}W^u(0)=W^s(0)$.
\item[H3] We shall make the additional technical assumption that a smooth co-ordinate change has been undertaken that flattens the stable and unstable manifolds within a
  small neighborhood $B$ of the origin. Moreover, we assume that within $B$ the
  dynamics is completely linear.
\end{description}
Note that the above hypothesis may be rigorously justified using $C^{1}$ conjugacy results for reversible systems,
see for example \cite{Sevryuk}. We omit the precise details here, because our purpose is to achieve 
a plausible argument for an asymptotic scaling, rather than a rigorous result.
We make a further assumption about the linearisation at the origin
\begin{description}
\item[H4] We suppose that at the parameter value $\eps=0$ the eigenvalues of
  the linearized  system within $B$ are degenerate and form two non-semisimple pairs
  of double real eigenvalues $\pm \rho$, for some $\rho>0$. Moreover we suppose
  that $\eps$ is a generic unfolding parameter that is designed
  in such a way that the characteristic polynomial
  of $Df(0;\eps,\nu)$ can be written 
\begin{equation}
  P(\lambda)= [(\lambda-\rho)^2-\eps ][(\lambda+\rho)^2-\eps] \:
  \label{eq:eig_assum} .
\end{equation}
Hence for $\eps<0$ the equilibrium is a saddle focus with eigenvalues 
$$
\lambda = \pm\rho\pm i\sqrt{-\eps} \: + \mathcal{O}(\eps), 
$$
whereas for $\eps>0$ there the origin is a real saddle with eigenvalues
$$
\lambda = \pm\rho\pm \sqrt{\eps} \: + \mathcal{O}(\eps), 
$$
see Figure~\ref{fig:pp} for corresponding phase portraits. 
\end{description}

Next, we make assumptions about the existence of a degenerate homoclinic orbit.
\begin{description}
\item[H5] We suppose that at the codimension-two point $\eps=\nu=0$
Are we also assuming that $\rho\ll \eps$) there exists a symmetric homoclinic orbit $y=\gamma(t)$ that is not wholly contained within $B$, and is such that 
$$
\gamma(0)\in S \quad \text{and}\quad \lim_{t\to\pm\infty} \gamma(t)=0.
$$
Moreover, we suppose that $B$ is chosen to be sufficiently close to the origin
that the trajectory $\gamma(t)$ contains only one connected component that is
outside of $B$. 
Thus we shall refer to $\gamma$ as being the {\em primary}
homoclinic orbit.  We also suppose that
$B$ is chosen such that the additional 
non-degeneracy condition \eqref{eq:nondeg0} is
satisfied. The particular degeneracy of the primary 
orbit we assume when $\nu=\eps=0$ is that there is a quadratic tangency between
  $W^u(0)$ and $S$ at $\gamma(0)$. We suppose that the parameter
  $\nu$ unfolds this tangency in a generic way such that, for all sufficiently small
  $|\eps|$, there are 
  two transverse intersections between $W^u(0)$ and $S$ near
  $\gamma(0)$ for small $\nu<0$, and no nearby intersection for small
  $\nu>0$ (see Figure~ \ref{fig:homofig} (b)).
  \end{description}
  \begin{description}
\item[H6] We assume certain non-degeneracy conditions concerning the general
  position of $W^u$ and $S$ at $\nu=0$, as set out in equations
  \eqref{eq:nondeg1} and \eqref{eq:nondeg2} in Section \ref{sec:glomap}
  below.
\end{description}

Our goal in what follows is to describe the fate of multi-pulse homoclinic
orbits in a neighborhood of $\gamma$ for small unfolding parameters $\eps$ and $\nu$ (assuming all other parameters
remain fixed). To do this
we use the well-established technique of approximation of the dynamics through
appropriate Poincar\'{e} maps (see Figure~\ref{fig:homofig}).

\subsection{Local map near the origin}
\label{sec:locm}
We shall now set up local co-ordinates within $B$. Given H1-H3
it is  possible to choose a local system of coordinates
$y=(y_1,y_2,y_3,y_4)^T$, such that he linear system can be written 
\begin{equation}
\label{eq:local}
\dot y=\mathbb{J} y\quad \text{where}\quad \mathbb{J}=\left(\begin{array}{cc}
 A_+  & 0\\
0 & A_-\end{array}\right),\quad \text{and}\quad  A_{\pm}=\left(\begin{array}{cc} \pm \rho & 1\\ \varepsilon &
  \pm \rho \end{array}\right).
\end{equation}
In this coordinate system, the dynamics in the local
stable and unstable manifolds 
$W^s_{\rm loc}(0)$ and $W^u_{\rm loc}(0)$ 
correspond to the uncoupled subsystems associated with matrices 
$A_-$ and $A_+$, respectively, as illustrated in Figure \ref{fig:pp}.
Under this construction, note that $y_2$ (and $y_4$) are in the direction of
the unstable (stable) eigenvector when $\eps=0$, and $y_1$ (and $y_3$) are in the direction of the generalized eigenvector. Moreover, without loss of generality we suppose that within the local co-ordinate system,  reversibility acts so that
$$
\mathcal{R} (y_1,y_2,y_3,y_4)^T \to (y_3,y_4,y_1,y_2).
$$
Furthermore, we choose the box $B$ to take the form
\begin{equation}
\label{eq:thebox}
B=\{y\in \mathbb{R}^4| y_1^2+y_2^2\leq h^2, \mbox{ and } y_3^2+y_4^2\leq h^2\}, \qquad \mbox{for some $h\ll 1$}.
\end{equation}
It is also useful to define polar coordinates within $B$
\begin{equation}
\label{eq:polars}
y=\left(\begin{array}{c}
          y_1\\y_2\\y_3\\y_4\end{array}\right)=\left(\begin{array}{c}
                                                       r_+\cos\theta_+\\r_+\sin\theta_+\\r_-\cos\theta_-\\r_-\sin\theta_-\end{array}\right),\quad
                                                   \text{with } 0\leq
                                                   r_{\pm}<h \text{
                                                     and } 0\leq \theta_{\pm} < 2\pi.  
\end{equation}

\begin{figure}
\begin{center}
\includegraphics[width=\textwidth]{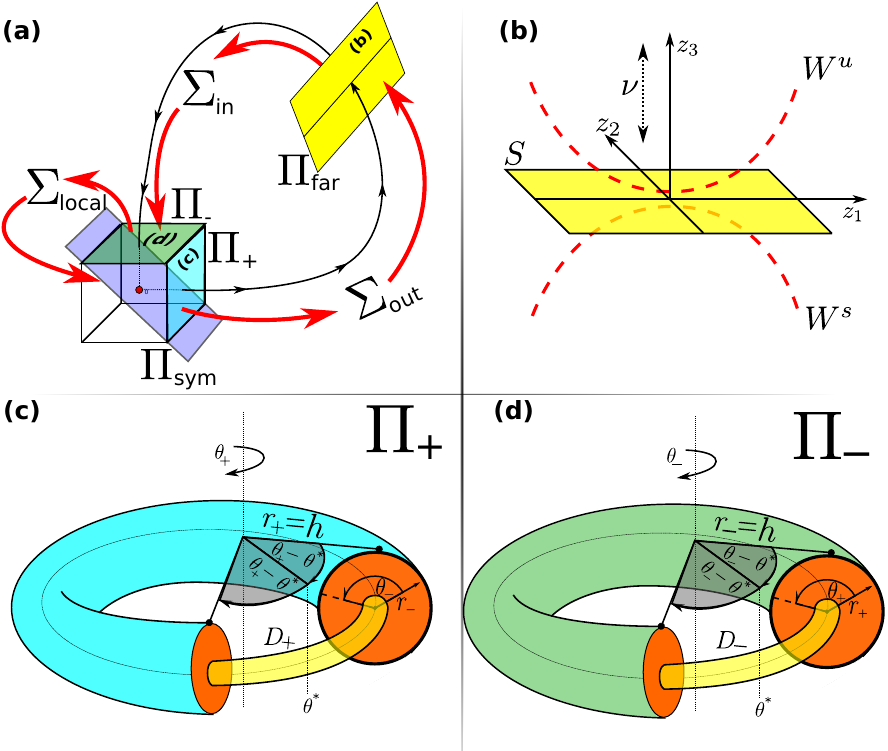}
\end{center}
\caption{Overview of the construction of the Poincar\'{e} maps in the
  vicinity of the codimension-two homoclinic orbit being analyzed,
  illustrating the geometry in each of the Poincar\'{e} sections
  $\Pi_+$, $\Pi_-$, $\Pi_{\rm sym}$ and $\Pi_{\rm far}$; see text for
  details.  (a) The primary homoclinic orbit is represented by a
  continuous black line, and the stationary point at the origin by a
  red dot.  The action of the Poincar\'{e} maps $\Sigma_{\rm out}$,
  $\Sigma_{\rm in}$, and $\Sigma_{\rm local}$ is indicated by bold red
  arrows.  Note the true dynamics are in $\mathbb{R}^4$, although the
  overall sketch is a 3D projection. Panels (b)-(d) indicate the
  three-dimensional geometry within each of the Poincar\'{e} sections
  $\Pi_{\rm far}$, $\Pi_+$ and $\Pi_-$, respectively.  In each case
  the local co-ordinate system is illustrated.}
\label{fig:homofig}
\end{figure}

We can now define local Poincar\'e sections on $\partial B$. In particular we
define incoming and outgoing sections 
\begin{align*}
\mbox{incoming:}&  \quad \Pi_-=\{y=(r_+,\theta_+,r_-,\theta_-)\in B \;|\; r_-=h\},
\\
\mbox{outgoing:} &  \quad \Pi_+=\{y=(r_+,\theta_+r_-,\theta_-)\in \mathbb{R}^4\;|\;r_+=h \}.
\end{align*}
It is straightforward to show that these sections are transverse to the flow, provided $|\eps|$ is sufficiently small. 
Another useful Poincar\'e section inside $B$ 
is the '{\em halfway through}' section, containing a local piece of $S$, given by
\begin{equation}
\label{eq:pis}
\mbox{halfway :}\quad  \Pi_{\rm sym}=\{y=(r_+,\theta_+,r_-,\theta_-)\in B\;|\;r_+=r_- \}.
\end{equation}
Note that, topologically, each of $\Pi_{{\rm sym},+,-}$ represents a
solid torus, see Figure~\ref{fig:homofig} (b) and (c) for an illustration.

Hypotheses (H5) and (H6) guarantee the existence of an
isolated homoclinic orbit $\gamma(t)$ when $\nu=\eta=0$. This orbit intersects
$\Pi_+$ at a single point 
$$\gamma_+ =\gamma\cap \Pi_+:\quad (r_+,\theta_+,r_-,\theta_-) = (h,\theta_*,0,0),$$
for some $\theta_* \in [0,2\pi)$. We shall assume
the non-degeneracy hypothesis 
\beq
\theta_* \neq 0 
\label{eq:nondeg0}
\eeq

By reversibility 
it intersects $\Pi_-$ at
$$
\gamma_-  =\gamma\cap \Pi_-: \quad (r_+,\theta_+,r_-,\theta_-) = (0,0,\theta_*,h).
$$  
We shall now define $D_\pm$ to be small neighborhoods of $\gamma_\pm$
in $\Pi_\pm$ (see panels (b) and (c) in  Figure~\ref{fig:homofig}):
$$D_- = \{ y \in \Pi_- \; | \; |\theta_- - \theta_*| \leq h, \; r_+
\leq h^2\},$$
$$D_+ = \{ y \in \Pi_+ \; | \; |\theta_+ - \theta_*| \leq h, \; r_- \leq h^2 \}.$$

In terms of polar co-ordinates, the linear equations can be written
\begin{align}
  \dot r_\pm &= r_\pm((1+\eps)\sin\theta_\pm \cos \theta_\pm  \pm \rho),
\label{eq:polar1}
  \\
\dot \theta_\pm &= \varepsilon\cos ^2\theta_\pm-\sin^2 \theta_\pm.
\label{eq:polar2}
\end{align}
We can obtain analytical solutions to the initial-value problem for \eqref{eq:polar1},\eqref{eq:polar2}. Note that the equations in the stable and unstable
subspaces are completely decoupled and can be solved independently. For definiteness we shall consider the dynamics in
the unstable subspace; the stable case
can be obtained by simply replacing $\rho$ by $-\rho$ in what follows. 
The equation for $\theta_+$ can be solved through separation
of variables, to obtain
\beq
\tan(\theta_+(t))=\sqrt{\varepsilon}\tanh(\sqrt{\varepsilon}
(t+C_{\theta_+}(\sqrt{\eps})),\quad
C_{\theta_+}(\sqrt{\eps})=\frac{\arctanh}{\sqrt{\varepsilon}}\left(\frac{\tan(\theta_+(0))}{\sqrt{\varepsilon}}\right),
\label{eq:theta_t}
\eeq
where $C_{\theta_+}$ is the constant of integration, to be can determined from the initial condition $\theta_+(0)=\theta_{+0}$ .
Substituting \eqref{eq:theta_t} into the $r_+$-equation, we can again
solve using separation of variables. The complete solution in terms
of the initial conditions 
$\left.(r_+,\theta_+)\right|_{t=0}=(r_{+0},\theta_{+0})$
can be written as
\begin{align}
\tan \theta_+(t) & =  
\frac{\varepsilon \tanh (\sqrt{\varepsilon} t)+\sqrt{\varepsilon}
      \tan\theta_{+0}}
      {\sqrt{\varepsilon}+\tanh(\sqrt{\varepsilon}
      t)\tan \theta_{+0} }
      \label{eq:the}\\
r_+(t) & =\sqrt{\frac{\Theta(t,\theta_{+0};\varepsilon)}{\Theta(0,\theta_{+0};\varepsilon)}}
  r_{+0}e^{\rho t},
\label{eq:r}
  \end{align}
where  
$$
\Theta(t;C_\theta;\varepsilon)
=(1+\varepsilon)\cosh(2\sqrt{\varepsilon}(t+C_{\theta}))+1-\varepsilon;
$$
that is,
\begin{align}
\Theta(t,\theta_{+,0};\varepsilon)
=&\frac{(1+\varepsilon)(\cosh(2\sqrt{\varepsilon}t)(\varepsilon+\tan^2\theta_{+,0})+2\sqrt{\varepsilon}\sinh(2\sqrt{\varepsilon}t)\tan\theta_{+0})}{(\varepsilon-\tan^2\theta_{+0})} +(1-\varepsilon). \label{eq:efe}
\end{align}

It will be useful to consider the limit $\varepsilon \to 0$ in the expressions (\ref{eq:the}) and (\ref{eq:r}). Using l'H\^opital's rule, the components of the solution are
\begin{equation}
\label{eq:theep0}
\lim_{\varepsilon\to 0} \theta_+(t) =\arctan\left( \frac{\tan \theta_{+0}}{1+t \tan\theta_{+0}}\right),
\end{equation}


\begin{align}
  \label{eq:simpli}
\lim_{\varepsilon\to 0}r_+(t)=r_{+0}e^{\rho t}\sqrt{(t \sin \theta_{+0}+\cos\theta_{+0})^2+\sin^2\theta_{+0}}.
\end{align}

Using these definitions, we can build a local map
$$
\Sigma_{\rm local}: \: D_-\subset\Pi_-\to \Pi_{\rm sym}.
$$
by following the local flow through the box $B$ until a trajectory
strikes $\Pi_{\rm sym}$; see Figure~\ref{fig:homofig}(a).

In terms of polar coordinates (\ref{eq:polars}), we have
$$\Sigma_{\rm  local}:\: \left(\begin{array}{c}  r_+^{(1)}\\  \theta_+^{(1)}\\  h\\  \theta_-^{(1)}\end{array}\right) \to \left(\begin{array}{c}  r^{(2)}\\  \theta_+^{(2)}\\  r_-^{(2)}\\  \theta_-^{(2)} \end{array}\right), \qquad \mbox{where} \quad r_+^{(2)}=r_-^{(2)}.$$
Using (\ref{eq:r}), we find
\begin{equation}
r_+^{(2)} =r_+^{(1)}\sqrt{\frac{\Theta(\tau,\theta_+^{(1)},\varepsilon)}{\Theta(0,\theta_+^{(1)},\varepsilon)}}e^{\rho\tau},\qquad 
r_-^{(2)} =h\sqrt{\frac{\Theta(\tau,\theta_-^{(1)},\varepsilon)}{\Theta(0,\theta_-^{(1)},\varepsilon)}}e^{-\rho
  \tau},
\label{eq:rpm}
\end{equation}
where $\Theta$ is given by (\ref{eq:efe}) and $\tau$ is the {\em time of flight}
from $\Pi_-$  to $\Pi_s$, which as yet unknown. The condition to be in
$\Pi_s$ is enforced by equating the two expressions in
\eqref{eq:rpm}. Thus, we obtain
\begin{equation}
\label{eq:tbox}
\left(\frac{r_+^{(1)}}{h}\right)^2
\frac{\Theta(\tau,\theta_+^{(1)},\varepsilon)
  \Theta(0,\theta_-^{(1)},\varepsilon)}{\Theta(\tau,\theta_-^{(1)},\varepsilon)
  \Theta(0,\theta_+^{(1)},\varepsilon)}=e^{-4\tau \rho}, \qquad
0<r_+^{(1)}<h \: . 
\end{equation}

Equation \eqref{eq:tbox} is a transcendental equation for $\tau$ in terms of the initial conditions
$r^{(1)}_+$, $\theta^{(1)}_+$ and $\theta^{(1)}_-$ in $\Pi_-$; 
in general, it  does not have a 
closed-form solution.
Nevertheless, note that $\tau=0$
when $r_+^{(1)}=h$. Conversely, when
$r_+^{(1)} \to 0$, which corresponds to approaching a solution in the stable manifold (which by construction never leaves $B$),   
the expression (\ref{eq:tbox}) reduces to
$e^{-4\tau\rho}=0$, implying $\tau \to \infty$ as expected. 
Note further, by definition of $D_-$, $r_+^{(1)}/h \ll 1$. This restriction implies that
we are interested in cases where $\tau\gg 1$. We are also interested in
the role of $\eps$ as an unfolding parameter, therefore it makes sense to consider the expression (\ref{eq:tbox}) asymptotically for 
large $\tau$  and in the limit
limit $\varepsilon\to 0$.
Using (\ref{eq:simpli}), we find in the limit $\varepsilon \to 0$, that
$$
\left(\frac{r_+^{(1)}}{h}\right)^2 \left(\frac{\sin^2\theta_+^{(1)}\tau^2
  +2\sin\theta_+^{(1)}\cos\theta_+^{(1)}\tau+1}{\sin^2\theta_-^{(1)}\tau^2
  +2\sin\theta_-^{(1)}\cos\theta_-^{(1)}\tau+1}\right)=e^{-4\tau \rho}.
$$
Now, for $\tau\gg 1$, the rational function on the left-hand side
can be approximated by its leading-order terms, from which we obtain
$$
\left|\frac{r_+^{(1)}}{h}\right|
\left|\frac{\sin\theta_+^{(1)}}{\sin\theta_-^{(1)}} + \mathcal{O}(\tau^{-1}) \right|=e^{-2\tau
  \rho}.
$$
Using this equation, an approximate expression for $\tau$ when $\eps$
and $r_+^{(1)}$ are small is
\begin{equation}
\tau=\frac{\log}{2\rho}\left(\left|  \frac{h\sin\theta_-^{(1)}}{r_+^{(1)}\sin\theta_+^{(1)}} + \mathcal{O}\left (\frac{1}{r_+^{(1)}} \right) \right |\right)
+ \mathcal{O}(\eps),
\label{eq:tau_approx} 
\end{equation}
which is valid for $\theta_-^{(1)} \neq n \pi \neq \theta_+^{(1)}$ for
any $n \in \mathbb{Z}$. This condition can be guaranteed by
redefinition of the origin of the angular co-ordinates if necessary.

Finally, substituting the expression \eqref{eq:tau_approx} for $\tau$
into the equations (\ref{eq:r}) and (\ref{eq:the}), we find an
explicit expression for the map $\Sigma_{\rm local}$ in the limit of
small $r_+^{(1)}$ and $\varepsilon$.
\bigskip

Using similar techniques, it is straightforward to use the estimates \eqref{eq:r} and \eqref{eq:simpli} to  
define the map that takes initial condition in
$D_-$ all the way through $B$ to $\Pi_+$ using similar techniques.
Such a construction is not necessary for the present analysis where we
only seek to study two-pulse orbits. But we would need that map if we
wanted to obtain a description of the complete dynamics near $\gamma$
or to construct three-pulse or higher order multi-pulse orbits (see
\cite{subsidiaryalan} in the non-degenerate case).

\subsection{Global Maps}
\label{sec:glomap}
Next we will define maps $\Sigma_{\rm out}$ 
(and $\Sigma_{\rm in}$) that
respectively flow in forward and backward time from a neighborhood
$D_+$ ($D_-$) along $\gamma(t)$ into a neighborhood of
$\gamma(0)$. Note that it is possible to construct such maps so that
they are images of each other under the reversibility
$\mathcal{R}$. In order to make such a construction, we need to define
a suitable local Poincar\'e section, $\Pi_{\rm far}$, say, transverse
to the flow at $\gamma(0)$; see Figure \ref{fig:homofig} (a).  In
order to build the maps, we need to define local coordinates
$(z_1,z_2,z_3)$ within $\Pi_{\rm far}$ as depicted in
Figure~\ref{fig:homofig} (b).

Note we are free to make this choice by defining $z_1$ to be the
co-ordinate direction that points along the direction of the tangency
between $W^u$ and $S$. Then we suppose that $z_2$ is the orthogonal
direction to this within $S$ and $z_3$ is the direction orthogonal to
$S$ within $\Pi_{\rm far}$. Under these assumptions, the reversibility
acts according to
\begin{equation}
\mathcal{R}\left(\begin{array}{c}z_1\\z_2\\z_3\end{array}\right)=\left(\begin{array}{c}z_1\\z_2\\-z_3\end{array}\right).
\label{eq:zr}
\end{equation}

Now, the map $\Sigma_{\rm out}$ maps $D_+\subset\Pi_+\to \Pi_{\rm
  far}$, our starting point $y^{(1)}$ lies in $D_+$.
Considering this limit in (\ref{eq:polars}), we find the initial point to be
$$
y^{(1)}=\left(\begin{array}{c} y_1\\y_2\\y_3\\y_4\end{array}\right)=\left(\begin{array}{c} h \cos(\theta^*+s) \\h\sin(\theta^*+s) 
\\r_-^{(1)}\cos\theta_-^{(1)}\\r_-^{(1)}\sin\theta_-^{(1)}\end{array}\right),
$$
where $s\in [-h,h]$ is a small parameter that parametrises the piece of $W^{u}_{\rm loc}$ in $D_+$, such that $s=0$ at the point $\gamma_+$. 
Consequently we can reduce the coordinates to $(s,y_3,y_4)\in \Pi_+$ and the map can be simplified to 
$$
\Sigma_{\rm out}:\left(\begin{array}{c}s\\y_3\\y_4\end{array}\right)\longmapsto\left(\begin{array}{c}
 z_1\\z_2\\z_3\end{array}\right). 
 $$
The map $\Sigma_{\rm out}$ maps $y^{(1)}$ into a point in
$\Pi_{\rm far}$, $s$ is a co-ordinate that parametrises a tangent
vector to trajectories in the unstable manifold at $\gamma_+$, and is
transverse to $\gamma(t)$. We expect such a vector to be mapped into
a direction tangential to the unstable manifold in $\Sigma_{\rm far}$
transverse to the primary homoclinic orbit.  Additionally, the
hypothesis H5 requires that this piece of $W^u \cap \Sigma_{\rm far}$
to be quadratically tangent to $S$ when $\nu=0$ (see Figure
\ref{fig:homofig}(b)). Notice too that the map should be a
diffeomorphism and hence invertible. Under these assumptions, and
assuming no further degeneracy, $\Sigma_{\rm out}$ can be written as
\begin{equation}
\label{eq:fmap}
\begin{pmatrix} z_1 \\ z_2 \\ z_3 
\end{pmatrix} 
=
\begin{pmatrix}
 as, \\ \eta_{23} y_3+\eta_{24}y_4 \\
 \eta_{33} y_3+\eta_{34}y_4+bs^2+\nu  
\end{pmatrix}  
+ \mathcal{O}(y^2),
\end{equation} 
for arbitrary real constants $a$, $b$ and $\eta_{i,j}$, satisfying
some mild hypotheses as described below.  The parameter $\nu$ controls
how the stable and unstable manifolds intersect with $S$ as specified
in H5 and depicted in Figure \ref{fig:homofig}(b). That is, for
$\nu>0$ the manifolds do not intersect there are is no primary
homoclinic orbit; whereas, for $\nu<0$, there are two transverse intersections. The sense of the tangency and the choice of co-ordinates implies that 
must assume that 
$$
a>0, \qquad b>0.
$$

The second global map, can be defined by exploiting the reversibility: 
$$
\Sigma_{\rm in}=\mathcal{R} (\Sigma_{\rm out})^{-1}\circ\mathcal{R},
$$
see Figure \ref{fig:homofig}.
The map $\Sigma_{\rm out}^{-1}:\Pi_{\rm far}\to \Pi_{+}$ is given by 
$$
\Sigma_{\rm out}^{-1}\left(\begin{array}{c}z_1\\z_2\\z_3\end{array}\right)=\left(\begin{array}{c}y_2\\y_3\\y_4\end{array}\right),\quad \text{where}\quad  s  =\frac{z_1}{a}, \quad y_3 =\frac{\left|\begin{array}{cc} z_2 &\eta_{24}\\\xi & \eta_{34}\end{array}\right|}{\psi},
\quad  y_4 =\frac{\left|\begin{array}{cc} \eta_{23} & z_2\\ \eta_{33} &                           \xi\end{array}\right|}{\psi},
$$ 
and $\xi=z_3-\nu -b\left(\frac{z_1}{a}\right)^2$. 
Notice that in order to be invertible we require, 
\beq
\psi:=\left|\begin{array}{cc} \eta_{23} & \eta_{24}\\
                            \eta_{33}  &
                                         \eta_{34} \end{array}\right| 
                                         \neq 0 .
\label{eq:nondeg1}
\eeq
In what follows, we shall also require further non-degeneracy assumptions;
namely that
\beq
\eta_{24} \neq 0 \qquad \mbox{and} \quad  0\neq \eta_{23} \neq  - \eta_{24} \tan(\theta^*).
\label{eq:nondeg2}
\eeq

Using the defined maps, we can establish the conditions for multi-pulse orbits.
We shall restrict attention to two-pulse orbits. In principle, $n$-pulse
orbits for any $n$ can be constructed similarly.  

\subsection{Constructing two-pulse orbits}
\label{sec:condbi}
 We start by considering a point $y_0$  near the primary orbit within $W^{u}$ in $D_+$ to subsequently find a set of conditions for the 
image of $y_0$ under the map 
$$
\Sigma_{\rm total}:D_+\longmapsto \Pi_{\rm sym}
$$ 
to lie in the intersection of the symmetric section with $\Pi_{\rm sym}$. By construction, such a point represents a homoclinic orbit
that passes a neighborhood of $\gamma(0)$ twice; a two-pulse orbit.

The map $\Sigma_{\rm total}$ can be defined as
\begin{equation}
\label{eq:mapacom}
\Sigma_{\rm total}=\Sigma_{\rm local}\circ\mathcal{R} \circ \Sigma_{\rm out}^{-1}\circ\mathcal{R}\circ
\Sigma_{\rm out}:D_{+}\longmapsto \Pi_{\rm sym}.
\end{equation}
We consider the initial point $y_0$ in a line segment 
of length $2s\ll h$, around the
primary homoclinic solution $\gamma$ within intersection of $D_+\cap
W^{u}(0)$. 
That is, we take
$$
r_-=0,\quad  r_+=h, \quad \theta_+=\theta^*+s, \quad |s|\ll 1,
$$
and consequently
\begin{equation}
\label{eq:yo}
y_0=\left(\begin{array}{c}y_1\\y_2\\y_3\\y_4\end{array}\right)=\left(\begin{array}{c}h\cos(\theta^*+s)\\h\sin(\theta^*+s)\\0\\0\end{array}\right).
\end{equation}
We proceed to compose each successive 
Poincar\'{e} map in
\eqref{eq:mapacom}.
The image of $y_0$ under the $\Sigma_{\rm out}$ is
$$
\Sigma_{\rm out}\left(\begin{array}{c} s\\0\\0\end{array}\right)\longmapsto\left(\begin{array}{c}as\\0\\bs^2+\nu\end{array}\right).
$$
Applying the reversing transformation \eqref{eq:zr}, we obtain
$$
\mathcal{R}\circ\Sigma_{\rm out}\left(\begin{array}{c} s\\0\\0\end{array}\right)\longmapsto\left(\begin{array}{c}as\\0\\-bs^2-\nu\end{array}\right),
$$
and after applying $\Sigma_{\rm out}^{-1}$, we obtain 
 $$
\Sigma_{\rm out}^{-1}\circ\mathcal{R}\circ\Sigma_{\rm out}\left(\begin{array}{c}s\\0\\0\end{array}\right)=\left(\begin{array}{c}s\\2\frac{\eta_{24}}{\psi}(\nu+bs^2)\\-2\frac{\eta_{23}}{\psi}(\nu+bs^2)\end{array}\right):=\left(\begin{array}{c}\zeta_2\\ \zeta_3
    \\ 
\zeta_4                       
   \end{array}\right)\in \Pi_+ \: . 
$$
We have to apply $\mathcal{R}$ again to reach $\Pi_-$. Within $B$ and its boundary, the reversing transformation acts to exchange
the roles of $\{y_1,y_2\}$ and $\{y_3,y_4\}$, and so we obtain
\begin{align*}
y_3& =h=r_-\cos\theta_- = r_- \cos(\theta* +\zeta_2) 
=r_-\cos(\theta^*+s) \\ 
y_4 & =r_-\sin\theta_- = r_- \sin(\theta* +\zeta_2) 
=r_-\sin(\theta^*+s) \\ 
y_1 & =\zeta_3 = \frac{2\eta_{24}}{\psi}(\nu+bs^2)=r_+\cos\theta_+, \\
y_2 & =\zeta_4=\frac{-2\eta_{23}}{\psi}(\nu+bs^2)=r_+\sin\theta_+. 
\end{align*}
After rearrangement, we find
\begin{align}
\label{eq:compo2}
r_-&=h,\\
\theta_-& =\theta^*+s,\nonumber\\
r_+&=\frac{2(|\nu+bs^2|)}{|\psi|}\sqrt{\eta_{24}^2+\eta_{23}^2},\\
\tan\theta_+&=-\frac{\eta_{23}}{\eta_{24}}=-\eta,\nonumber\\
  \sin\theta_+&=\frac{-\eta_{23}}{\sqrt{\eta_{23}^2+\eta_{24}^2}}\quad \text{which implies}\quad 
              r_+\sin\theta_+=-2\eta_{23}\left|\frac{\nu+bs^2}{\psi}\right|=-\frac{1}{\varphi} |\nu+bs^2| , \nonumber
\end{align}
where $\eta$ and $\varphi$ have been defined to simplify the subsequent algebra. Note that 
$\varphi$ is well defined and non-zero by the 
nondegeneracy assumption \eqref{eq:nondeg2}

 We finally need to apply the map $$
\Sigma_{\rm local}: \left(\begin{array}{c} r_+^{(1)}\cos\theta_+^{(1)}\\
                  r_+^{(1)}\sin\theta_+^{(1)}\\
                  r_-^{(1)}\cos\theta_-^{(1)}\\
                  r_-^{(1)}\sin\theta_-^{(1)}\end{array}\right)\longmapsto
 \left(\begin{array}{c} r_+^{(2)}\cos\theta_+^{(2)}\\
                  r_+^{(2)}\sin\theta_+^{(2)}\\
                  r_-^{(2)}\cos\theta_-^{(2)}\\
                  r_-^{(2)}\sin\theta_-^{(2)}\end{array}\right), 
$$ 
and impose that the image is in the symmetric section. 
This is the condition for the existence of \textsl{two-pulse} orbits. In terms of our coordinates, we need that 
\beq
r_+^{(2)}=r_-^{(2)}\quad \text{and} \quad
\theta_+^{(2)}=\theta_-^{(2)}.
\label{eq:final_conds}
\eeq
The first condition is automatically met by the definition of the
Poincar\'e section $\Pi_{s}$ (\ref{eq:pis}) and also by definition of
the time  of flight $\tau$. Hence, it is sufficient to
impose the second condition to establish the existence of the a two-pulse homoclinic orbit. 
Using (\ref{eq:the}), we obtain 
$$
\tan\theta_+^{(2)}=\frac{\varepsilon \tanh
      (\sqrt{\varepsilon} \tau)+\sqrt{\varepsilon}
      \tan\theta_+^{(1)}}{\sqrt{\varepsilon}+\tanh(\sqrt{\varepsilon}
      \tau)\tan \theta_+^{(1)}}=\frac{\varepsilon \tanh
      (\sqrt{\varepsilon} \tau)+\sqrt{\varepsilon}
      \tan\theta_-^{(1)}}{\sqrt{\varepsilon}+\tanh(\sqrt{\varepsilon}
      \tau)\tan \theta_-^{(1)}}=\tan\theta_-^{(2)}.
    $$
        That is, we are looking for
\beq
\frac{\varepsilon
  A+\sqrt{|\varepsilon|}b}{\sqrt{|\varepsilon|}+AB}=\frac{\varepsilon
  A+\sqrt{|\varepsilon|}c}{\sqrt{|\varepsilon|}+AC},
\label{eq:simpcond}
\eeq
with
$A=\tanh(\sqrt{\varepsilon}\tau)$ if $\varepsilon>0$ or 
$A=\tan(\sqrt{-\varepsilon}\tau)$ if $\varepsilon<0$, 
$B=\tan\theta_+^{(1)}$, 
and $C=\tan\theta_-^{(1)}$. 
The condition \eqref{eq:simpcond} will be true satisfied
whenever 
$$
A^2=1 \quad  \text{or}  \quad B=C.
$$
Rewriting $A$, $B$ and $C$ in terms of the expression of
(\ref{eq:compo2}), we note that, 
irrespective of the sign of $\eps$, $B=C$ would correspond to 
$\tan \theta_-^{(1)}=\tan (\theta^*+s)\approx \tan \theta^*+s\sec^2\theta^*=-\eta=\tan \theta_+^{(1)}$. But, given that $s$ is a small parameter, this condition is ruled out
by the nondegeneracy assumption \eqref{eq:nondeg2}.
We are left with the case $A^2=1$, the number of 
solutions to which will be qualitatively different depending on
the sign of $\varepsilon$. That is, we seek solutions
\beq
\tanh(\sqrt{\varepsilon}\tau)= 1, 
\label{eq:tanhcond}
\eeq
and we need to consider seperately the cases 
$\eps>0$ and $\eps<0$. 

Taking first the case $\eps>0$, restricting to positive values of time $\tau$ and using the approximation
(\ref{eq:tau_approx}), from the condition \eqref{eq:tanhcond} we obtain
\beq
\left.\tau\right|_{\varepsilon=0}
\approx \frac{\log}{2\rho}\left(\left|\frac{h\varphi \sin(\theta^*) +\mathcal{O}(s) }{ \nu+bs^2}\right|\right)= + \infty,\quad \text{which implies}\quad \nu+bs^2=0. 
\label{eq:exist1}
\eeq
Given that $b>0$ by assumption, we find this equation predicts 
precisely two solutions when $\nu<0$, namely
$s=\sqrt{|\nu|/b}$, which correspond to the 
two \emph{primary} homoclinic orbits. There are no two-pulse orbits in this case, and there are no solutions at all for $\nu>0$. 

In contrast, for $\varepsilon <0$, the condition \eqref{eq:tanhcond} becomes $\tan(\sqrt{|\varepsilon|}\tau)= 1$. Solving this equation after using the approximation (\ref{eq:tau_approx}) we obtain
\beq
\sqrt{|\varepsilon|}\left(\left.\tau\right|_{\varepsilon=0}\right) \approx \sqrt{|\varepsilon|}\frac{\log}{2\rho}\left(\left|\frac{h\varphi \sin(\theta^*) }{
    \nu+bs^2}\right|\right) = \frac{\pi}{4}+n\pi    \qquad n \in \mathbb{Z}_+ ,
\label{eq:tan_formula}
\eeq
where we have used the fact that 
$\sin(\theta^*) \neq 0$ and $s \ll 1 $ is small. Here $n$ counts the number of
half-oscillations close to $W^s$ within the box $B$, before hitting the symmetry
condition. 

Recalling that $b>0$ and $\varphi\neq 0$, 
we can see that for each small $\nu<0$ and $n$
sufficiently large there will be two infinite 
sequences $s^{\pm}_n$ of $s$-values that solve 
\eqref{eq:tan_formula}. Each of these $s$-values corresponds to the existence
of a two-pulse orbit. Moreover these sequences  
converge to $\pm \sqrt{(\nu/b)}$ 
as $n\to \infty$, which are the $s$-values of the two primary homoclinic
orbits. 
That there is a family of two-pulses converging on each primary orbit is 
consistent with the result in 
\cite{subsidiaryalan} 
for the case $\nu<0$ and
$|\varepsilon|=\mathcal{O}(1)$. 

Finally, consider solutions to \eqref{eq:tan_formula} for 
$\varepsilon<0$ and $\nu>0$. Now note that for each $\nu$ there will finitely
many pairs of $s$-values leading a solution, up to some $n_{\rm max}(\mu)$. 
Moreover, there are a sequence of $\nu$-values converging on $\nu=0$ at which $n_{\rm max}$ increases by one, due to a double route of 
\eqref{eq:tan_formula}. Thus as $\nu$ increases from zero, pairs of two-pulse
orbits are destroyed in fold bifurcations. 

\section{Asymptotic predictions of the analysis}
\label{sec:asymptotics}

A careful analysis of the existence of primary and two-pulse orbits according to \eqref{eq:tan_formula} and \eqref{eq:exist1} for small $\eps$ and $\nu$
leads to a two-parameter bifurcation
diagram that is 
summarized qualitatively in 
Fig.~\ref{fig:epsnu}. In addition, 
we can make specific quantitative predictions,
that can compared
with numerical results for the two example
systems presented in Sec.\ref{sec:examplesss}. 

\subsection{Scaling laws}

\begin{figure}
\begin{center}
  \includegraphics[width=1\textwidth]{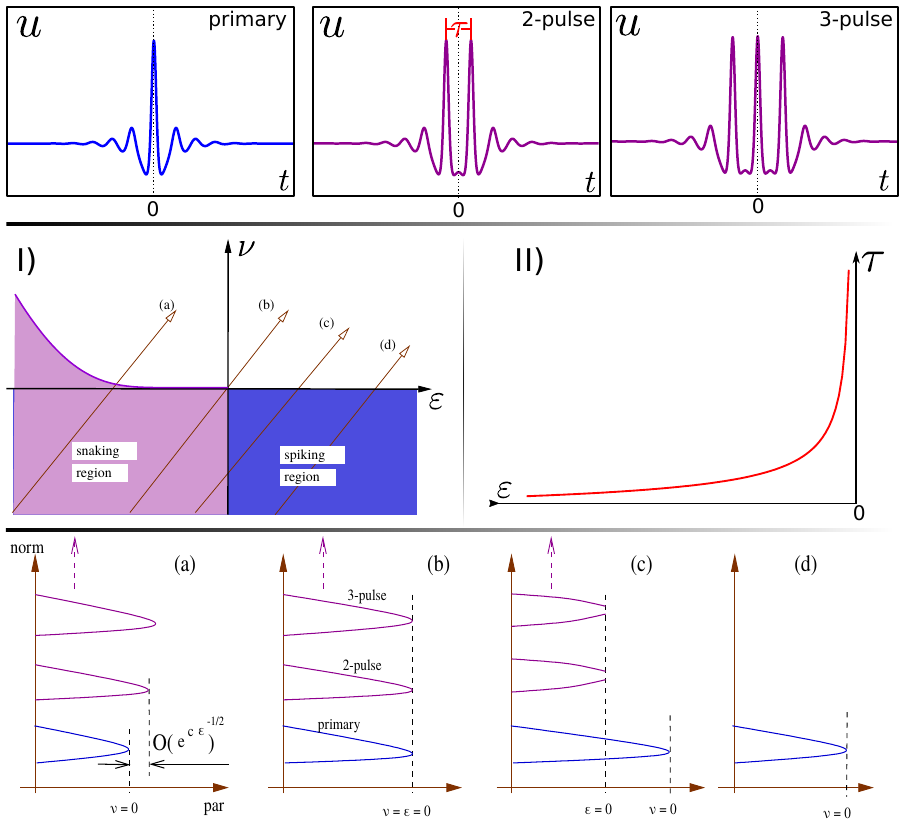}
\end{center}
\caption{Qualitative summary of the results on bifurcation analysis of
primary (blue) and multi-pulse (pink) homoclinic orbits. \textsl{Top panel} close to the codimension-two point 
$\eps=\nu=0$. The \textsl{middle panel} shows the two main results:
(I) the time of flight (distance peak to peak) in the 2-pulse
solutions; (II) The region of existence of multi-pulse solution
(snaking region), the sub-figures indicate the results of one-parameter continuation
along each of the depicted paths (a)--(c).}
\label{fig:epsnu}
\end{figure}

First, for 
$\varepsilon<0$ the analysis predicts the existence of infinite
two-pulse orbits. We can estimate the 'time of flight' between the two humps for small $\eps$ by considering the left-hand side of (\ref{eq:tan_formula}). 
Specifically, the time spent inside the box $B$ is
\begin{equation}
  \tau \approx \frac{\pi}{\sqrt{|\varepsilon|}}\left(\frac{1}{4}+n\right)\propto \frac{1}{\sqrt{|\varepsilon|}}
    \label{eq:pred1}
\end{equation}
where the positive integer $n$ counts the number
half-oscillations in $\theta_-$ close to 
$W^{s}$ before the symmetry point of the 2-pulse orbit.
In the limit of $\eps \to 0$, 
this time $\tau$ is an approximation (up to an $\eps$-independent constant) of 
the time-interval between the two approximate copies
of the primary orbit that are concatenated to form
the two-pulse orbit. 
(see the mid panel in the
top of Figure \ref{fig:epsnu} in red).  
Consequently, for small $\nu$ and $\eps$,
(\ref{eq:pred1}) predicts that the distance between maxima in the graphs of two-pulse orbits 
will increase like the reciprocal of as $\sqrt{|\varepsilon|}$ 
as $\eps$ tends to zero (see
the panel II in Figure \ref{fig:epsnu}).

Second, the analysis predicts the location of folds of
two-pulse orbits. Assuming that $\eps<0$, and solving for $\nu$, from expression (\ref{eq:tan_formula}) we obtain an
expression for the $\nu$-dependence of a two-pulse orbit
as a function of the internal parameter $s$:
    \begin{equation}
      \nu^{(n)}(s) =  h\varphi \sin(\theta^*) \exp \left(-\frac{2\rho \pi}{\sqrt{|\varepsilon|}}\left(\frac{1}{4}+n \right) \right)   -bs^2 .
      \label{eq:nube}
      \end{equation}
Notice that this expression for $\nu$ has a quadratic dependence on the internal parameter $s$ with an extremum at
$s=0$. This gives the existence of a fold two-pulse 
homoclinic orbits with respect to $\mu$:
      \begin{equation}
        \nu^{(n)}_{\rm fold}= h\varphi \sin(\theta^*) \exp\left ( \frac{-2\rho \pi}{\sqrt{|\varepsilon|}}   \left(n+\frac{1}{4}\right) \right ) \: \propto \exp \left (\frac{-2\rho\pi}{\sqrt{|\varepsilon|}} \right), \quad n\in \mathbb{Z_+},\quad \varepsilon <0.
          \label{eq:bimfold}
        \end{equation}
For each $n$ we can regard \eqref{eq:bimfold} as providing
a curve in the $(\varepsilon,\nu)$ parameter plane in which two-pulse orbits get destroyed in a fold  bifurcations. A version of this curve for a fixed
$n$ is illustrated in Figure \ref{fig:epsnu}(I). 
Notice that curve (\ref{eq:bimfold})
emanates from the codimension-two
point $(\eps,\nu)=(0,0)$. Thus two-pulse orbits are destroyed in a fold 
for positive $\nu$-values which, as 
$\eps \to 0$, tends exponentially quickly to the 
same parameter value $\nu=0$ as the primary fold.  
It is in principle a straightforward
argument to show that all $n$-pulse for $n>2$ should
also be destroyed in this manner as $\nu$-increases
(see path (a) in  Figure.~\ref{fig:epsnu} panel I) and the respective 
one-parameter continuation sketch in the bottom panel).

We can now ask the question what happens to the same orbits as we move from path (a) to path (c) in Fig.~\ref{fig:epsnu} via
path (b) which passes through the codimension-two point. 
On path (c)
the multi-pulse orbits first cross $\eps=0$ before reaching $\nu=0$ and therefore each multi-pulse orbit is destroyed via a bifurcation at which $\tau \to \infty$.
This is a bifurcation in which the large pulse-like maxima become 
infinitely far seperated. 

We should stress that there is no guarantee from this analysis that the two-pulse orbit that are
connected globally to the primary via the homoclinic snake should be 
destroyed via this mechanism, it is just that all multi-pulse orbits 
that pass in the vicinity of the primary homoclinic orbit
  $\gamma$ inside the box B must be destroyed in this
  manner. Nevertheless, we note from the 
  numerical results in the Sec.\ref{sec:examplesss} indicate that in both of the examples, the multi-pulse orbits within the homoclinic
snake do indeed get destroyed at the Belyakov-Devaney transition via the mechanism
we have just outlined. In
particular, the bottom panel of Fig.~\ref{fig:pmap_wp} (see inset (c)),
shows the one-parameter continuation of
the two-pulse orbit. In that case, the bifurcation diagram exhibits the
same behavior predicted by this analysis (cf.~Fig.~\ref{fig:epsnu}(c)),
where the two-pulse homoclinic orbit
terminates precisely at the Belyakov-Devaney transition.

\subsection{Comparison with numerical results}

In order to conclude the analysis, we shall compare the
above quantitative predictions with 
numerical results obtained from the two example systems studied in
Sec.~\ref{sec:examplesss}. 
More precisely, we shall
seek evidence for the scaling predictions 
(\ref{eq:pred1}) for the pulse-to-pulse seperation $\tau$, and
(\ref{eq:bimfold})) for the the $\nu$-value of the fold
as a function of $\eps$. 

We should stress at the outset that
there extreme numerical difficulties with attempting such verification. As has been documented elsewhere, e.g 
\cite{Oldeman,yuvalss}, it is a difficult numerical task to accurately compute a multi-pulse homoclinic in the limit (equivalent to $\eps \to 0^-$ here) that the pulses become infinitely far seperated. The inherent problem is that two graphs of the primary orbit placed side by side with arbitrary seperation $\tau$ approximately solve the system up to an error that is exponentially small as $\tau \to \infty$. Hence 
any initial-value or boundary-value solver with finite
precision is unable to distinguish true two-pulse orbits.
For this reason, we have been unable to trace two-pulse 
orbits particularly far into the limit $\eps\to 0^-$. 

In each of the two examples, our numerical procedure was
as follows. 
Using two-parameter continuation in Auto \cite{auto}, we find
accurately locate the codimension-two point at which there is a fold of the primary homoclinic orbit with respect to a parameter and simultaneous the imaginary part of the eigenvalues of the equilibrium vanish. We then compute 
the two-pulse orbit that is globally part of the homoclinic 
snake and follow the fold of this orbit in the same
two parameters. Along this curve we monitor the eigenvalues of
the equilibrium and define a parameter $\eps$ as in the 
analysis such that the imaginary part of the eigenvalues
are $\pm \sqrt{-\eps}$. We then define a parameter that is equivalent to $\nu$ by projecting the displacement in the parameter plane to 
the curve on which the primary orbit has its fold, onto the 
the direction to curves $\eps=$const.~to be equivalent to $\nu$ from the analysis.
The origin of $\nu$
In this manner we probe the region of the that is distinguished with a maroon star in the top panel of Figures
\ref{fig:pmap_wp} and \ref{fig:political_crime}, measuring as we do the distance between the two large peaks (equivalent to
$\tau$ up to a constant). 

\begin{figure}
\begin{center}
\includegraphics[width=\textwidth]{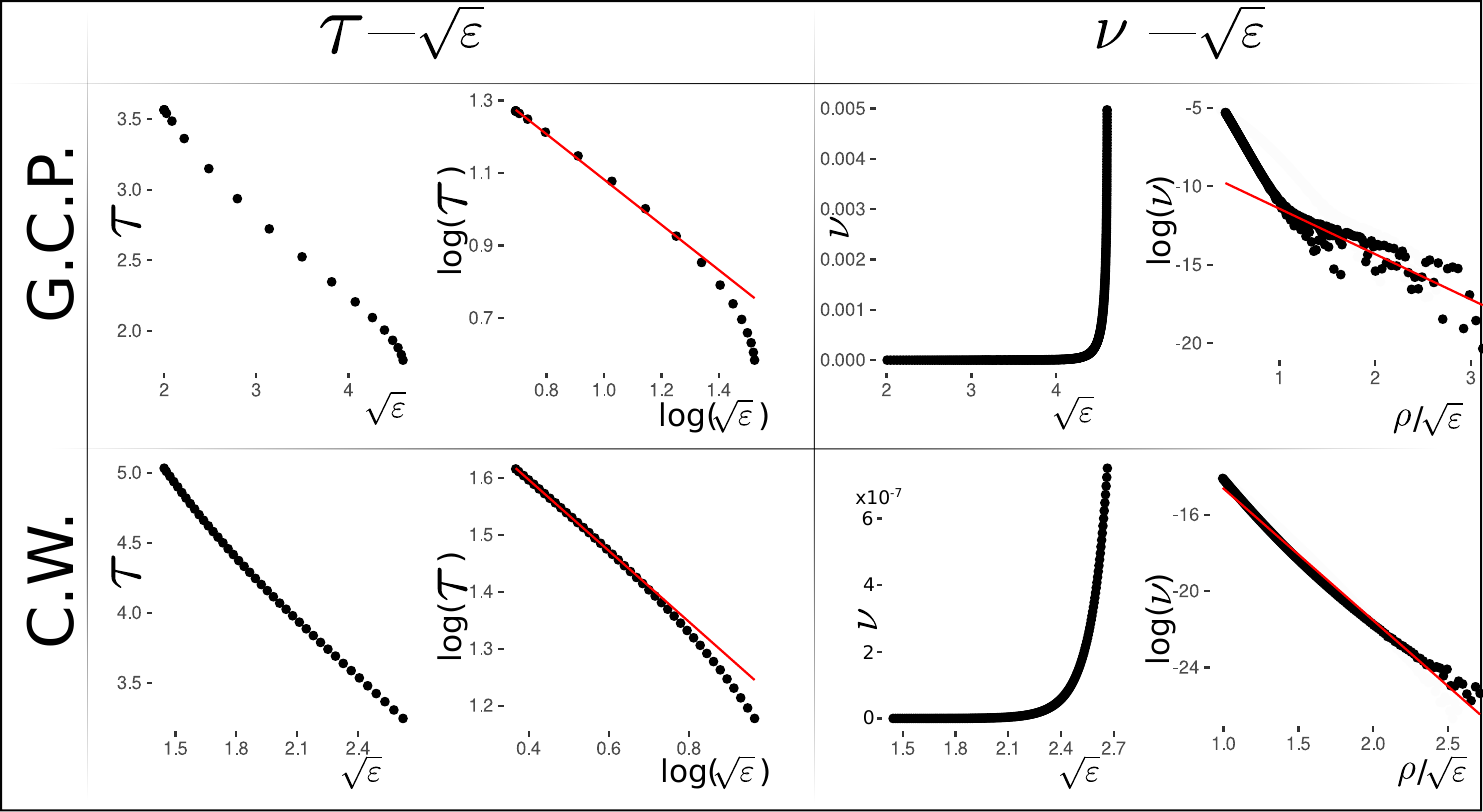}
\end{center}
\caption{Comparing between the asymptotic predictions for $\tau$
  (\ref{eq:pred1}) (\textsl{Left side}) and $\nu$ 
  (\ref{eq:bimfold}) (\textsl{Right side}) as a function of the
  eigenvalues and results of the numerical continuation. The
  \textsl{Top} (\textsl{Bottom}) half illustrates these results in the (G)eneralized (C)ell
  (P)olarity Model ((C)rime (W)ave model). Black points correspond to
  the results of the numerical continuation and the red lines
  represents the linear fit. See text for details.}

\label{fig:compa}
\end{figure}

The results for both models 
are summarized in Fig.~\ref{fig:compa}.
In both cases, for the numerical reasons outlined above, we find that the curve of folds of the two-pulse orbit fails to converge at some significant distance
from the codimension-two point, equivalent to $\eps=\mathcal{O}(1)$.  

The results for the pulse-to-pulse distance as a function of
$\sqrt{|\varepsilon|}$ are presented in the left-hand panels of the Fig.~\ref{fig:compa}. Not how, in both models, the time of flight $\tau$ increases algebraically as $\varepsilon$ tends to zero. Furthermore, considering the
logarithm of (\ref{eq:pred1}), we obtain
\begin{equation}
  \log\tau= -c_1 \log \sqrt{\eps}+ c_2, \quad \text{ with }
c_1=1 \quad \mbox{and } c_2=\log \left[\pi\left(n+\frac{1}{4}\right) \right ],
\label{eq:predlintau}
\end{equation}
suggesting that, under this transformation, a linear model could be fit to the numerical results. 
Such a linear fit is indicated
in the next right-most panels of the figure. 

The third column of pabels in Fig.~\ref{fig:compa} shows 
numerical results for location of the $\nu$-value distance between the fold of the two-pulse and primary orbit.
The exponential-like behavior
exhibited by both models is apparent. 
By considering the logarithm of expression
(\ref{eq:bimfold}) one can obtain a straight line prediction
on a log-log plot:
\begin{equation}
\log \nu_{\rm fold}=-c_3  \frac{\rho}{\sqrt{\eps}} + c_4, \quad \mbox{
  with } c_3 = 2\pi\left(n+\frac{1}{4}\right), \quad c_4 = \log(h\varphi \sin(\theta^*) ).
\label{eq:predlinu}
\end{equation}
The final column shows that such a linear prediction can be observed in the limit of small $\sqrt{\eps}$.

While these results are encouraging, there is little evidence
that the constants $c_1$--$c_4$ match those of the theory. 
The constants found in the least-squares linear regression fits in
Fig.~\ref{fig:compa} are given in Table \ref{tab:nr}.
Note that 
As $\eps \to 0$, in addition to observing the expected behaviors
(i.e. $\tau\to \infty$ and $\nu \to 0$), we also can also see the
transition between non-linear (large $\eps$) to linear (smaller
$\eps$) regimes. These changes are more pronounced in the log and
semi-log scales (cf. Figure \ref{fig:compa}). Since we are testing
(\ref{eq:predlintau}) and (\ref{eq:predlinu}), we therefore used a sub-interval of $\eps$ and $\rho/\eps$, respectively. The rightmost column in Tables
\ref{tab:nr} (a) and (b) specifies the range used in each case.

\begin{table}
\begin{center}
  \begin{tabular}{ll}
    (a) Prediction $\tau(\sqrt{\eps})$ & (b) Prediction $\nu(\sqrt{\eps},\rho)$\\
      \begin{tabular}{|c|c|c|c|c|}\hline
        Model & $c_1$ & $c_2$ & $R^2$ & Range\\\hline
        G.C.P. & 0.62 & 1.71 & 0.997 &$\log \sqrt{|\eps|}<1.4$\\ 
        C.W. & 0.62 & 1.84 & 0.999 &$\log \sqrt{|\eps|}<0.65$\\ \hline
      \end{tabular}

      &

        \begin{tabular}{|c|c|c|c|c|}\hline
        Model & $c_3$ & $c_4$ & $R^2$ & 
        Range \\\hline
        G.C.P. & 2.89 & 8.53 & 0.826 &$\frac{\rho}{\sqrt{\eps}}>$1\\ 
        C.W. & 6.89 & 7.73 & 0.990 &$\frac{\rho}{\sqrt{\eps}}>0.1$\\ \hline
      \end{tabular}
  \end{tabular}
\end{center}
\vspace{.1cm}
\caption{Linear regression of the numerical continuation results using
  the linear models for $\tau$ (\ref{eq:predlintau}) (a) and
  $\nu$  (\ref{eq:predlinu}) (b) respectively. The result of
  these linear regression have been illustrated in Figure \ref{fig:compa}.}
  \label{tab:nr}
  \end{table}
The linear regression confirms that in each case we observe a linear
behavior (see the $R^2$ values in Table \ref{tab:nr}). Additionally,
the approximation for $c_1$ and $c_2$ in Table \ref{tab:nr}(a)
are remarkably similar between models. However, the match to
the theoretical value $c_1=1$ is poor, presumably because 
$\sqrt{\eps}$ is still a long way from the limit $\eps=0$
where the approximation is valid. 
In contrast, the $\nu-\eps$
linear fit presents differences between $c_3$ and $c_4$.


\section{Discussion}
\label{sec:discussion}
In this paper we have reported progress in the description of the
transition between spike-like and localized patterned states in
systems of reaction diffusion equations. That the scenario have been observed in two separate examples, suggest that this is a {\em generic} situation.

The key we have found to understanding the transition in both
examples is that, in a two-parameter space, the localized-structure
region is bisected by a line corresponding to the codimension-one
Belyakov-Devaney transition.  To that end we provided a partial
unfolding of the codimension-two degeneracy where a fold of a primary
homoclinic orbit coincides with the Belyakov-Devaney transition.  By
introducing two parameters, $\eps$ controlling the linearization and
$\nu$ controlling the primary homoclinic fold, we find the
asymptotic scalings of how all subsidiary multi-pulse orbits are
destroyed in a neighborhood in parameter space. 

Our main contribution has been to find approximate expressions for
the local and global Poincar\'{e} maps in a neigborhood of the degenerate
homoclinic orbit we study. This is particularly challenging because of the lack of a closed form analytical expression for the local map, and because of the need to unfold the quadratic tangency in the global map in a generic way. We
are unaware of any previous work that has attempted to unfold such a 
codimension-two singularity. 

The main
results of the analysis are summarized qualitatively in Fig.~\ref{fig:epsnu}.
In particular, according to \eqref{eq:bimfold}, we find that the $\nu$-values of the
folds of all the subsidiary orbits become exponentially close to that
of the primary orbit as $\eps \to 0$. This prediction from the
analysis, and the prediction \eqref{eq:pred1} on the rate of
separation multi-pulses as $\eps \to 0$ for $\nu<0$, were tested
against two example systems and found to be consistent. Precise 
quantitative verification is problematic though due to inherent
numerical difficulties in computing multi-pulse
orbits close to the codimension-two point, 
as has been report elsewhere in this
contex (see e.g.~\cite{Oldeman,yuvalss}). Perhaps computation using arbitrary
precision arithmetic might resolve this problem, but that is beyond the scope
of the present work. 

We should also point out that what we have found is not the only way
that a homoclinic snake can become more complex than the simple
picture in Figure.~\ref{fig:double_snake}.  A number of generic
possibilities of the structure of homoclinic curves within the pinning
region are analyzed in \cite{Makrides}. In essence what we have
analyzed is a situation in which the left-hand fold of the snake in
Figure.~\ref{fig:double_snake} crashes into another bifurcation. Another
possibility is that the right-hand fold of the snake should approach
the parameter value of the Turing bifurcation, which has been seen in
a number of systems and also leads to $C$-shaped isolas. In contrast
though all homoclinic orbits grow algebraic rather than exponential
tails in this limit, rather than multi-pulses disappearing to infinity.
Another possibility is that the left-hand folds of the snake coincide
with a fold of the primary pattern forming periodic orbit. This
situation was analyzed in \cite{ChampWagen} and leads to $S$-shaped
isolas in which compound small and large pulses form {\em defect
  mediated snaking} \cite{Defect}.

It should also be stressed that all we have shown is the mechanism by
which infinitely many multi-pulse homoclinic orbits are destroyed as a
fold of a primary homoclinic orbit reaches a Belyakov-Devaney
transition. However, there is no {\em a priori} reason that these orbits should be
part of what was the homoclinic snake born at the codimension-two
degenerate Turing bifurcation (the red circle in
Figs.~\ref{fig:pmap_wp} and \ref{fig:political_crime}).  In fact, the analysis of homoclinic snaking
\cite{alansnake,snakebeck,Makrides} occurs due to an unfolding of a
heteroclinic orbit between a hyperbolic equilibrium and a generic
saddle-like periodic orbit. In principle, as the equilibrium passes
through a double-real-eigenvalue transition it remains hyperbolic and
the dimensions of the stable and unstable manifolds in question
do not change.  It just seems that for the particular examples we
have studied, the periodic orbit that is involved in the heteroclinic
connection becomes the smaller-amplitude homoclinic orbit passing
through the Belyakov-Devaney point. (See for example the bottom Figure.~\ref{fig:pmap_wp}(c) in
which the multi-pulse orbit being destroyed in the upper panel is
actually a hybrid of the large-amplitude and the small-amplitude
homoclinic orbit).  Why this periodic orbit should become homoclinic
is not clear. The question would seem to be tied up with an unfolding
of the singular point in which there is a quadruple zero eigenvalue
equilibrium in which the diffusion ratio is also zero (equivalent to
$\gamma=\delta=0$ in our first example).  

As further evidence of the
ubiquity of what we have found here, forthcoming work will show that
the same transition between spikes and localized patterns as analyzed
here also occurs in a wide variety of Schnakenberg-type models in the
presence of symmetry-breaking terms \cite{Fahad}.  We also mention the
related work \cite{parra2} on the Lugiato-Lefever equation which has a
related transition to the one here, but for which the fundamental
codimension-two bifurcation has a different structure.

There remain many aspects of this problem that we have not fully
analyzed, in addition to an unfolding of this singular codimension-two point.  For example, there is more that can be said about the ordering of
two-, three-, etc.-~pulse orbits and prediction that the disappearance of
multi-pulse orbits is via hybrids between the large and
small-amplitude homoclinic orbits. The fact that the $W^{u}(0)$ is
two-dimensional means that homoclinic orbits have a specific ordering
and disappearance of certain orbits through a fold has implications on
non-existence of certain other orbits at that parameter value.
Extensive arguments of this nature were previously used in
\cite{Buffoni} in a related context.

\section*{Acknowledgments}
The authors acknowledge helpful conversations with Pedro Parra-Rivas,
Damia Gomila, Lendert Gelens, Edgar Knobloch, Marcel Clerc, Jens Rademacher and Yuval 
Zelnik.  Nicolas Verschueren would like to acknowledge  
``Programa de doctorado en el Extranjero Becas Chile Contract No.72130186''
for PhD funding.

\ifsiam
\bibliographystyle{siamplain}
\else
\bibliographystyle{plain}
\fi
\bibliography{biblio}

\begin{thebibliography}{10}

\bibitem{revoptls}
{\sc T.~Ackemenn and W.~Firth}, {\em Chapter 6: Fundamentals and applications
  of spatial dissipative solitons in photonic devices}, in Advances in Atomic
  Molecular and Optical Physics, vol.~57, Academic Press, 2009, pp.~323-- 421.

\bibitem{Akhmediev}
{\sc N.~Akhmediev and A.~Ankiewicz}, {\em Dissipative Solitons},
  Springer-Verlag, Berlin, 2005.
\newblock Lecture Notes in Physics.

\bibitem{Fahad}
{\sc F.~{Al Saadi}, A.~Champneys, and N.~Verschueren}, {\em Universal structure
  of localized patterns in {S}chnackenberg-like models}, 2020.
\newblock In preparation.

\bibitem{Sandstede}
{\sc J.~Alexander, M.~Grillakis, C.~Jones, and B.~Sandstede}, {\em Stability of
  pulses on optical fibers with phase-sensitive amplifiers}, Z. Angew. Math.
  Phys., 48 (1997), pp.~175--192.

\bibitem{Lloyd2}
{\sc D.~Avitabile, D.~Lloyd, J.~Burke, E.~Knobloch, and B.~Sandstede}, {\em To
  snake or not to snake in the planar {S}wift-{H}ohenberg equation}, SIAM J.
  App. Dynamical Sys., 9 (2010), pp.~704--733.

\bibitem{barafamous}
{\sc I.~V. Barashenkov, M.~M. Bogdan, and V.~I. Korobov}, {\em Stability
  diagram of the phase-locked solitons in the parametrically driven, damped
  nonlinear schr{\"{o}}dinger equation}, EPL, 15 (1991), p.~113.

\bibitem{brara2}
{\sc I.~V. Barashenkov, E.~V. Zemlyanaya, and T.~C. van Heerden}, {\em
  Time-periodic solitons in a damped-driven nonlinear schr\"odinger equation},
  Phys. Rev. E, 83 (2011), p.~056609.

\bibitem{snakebeck}
{\sc M.~Beck, J.~Knobloch, D.~Lloyd, B.~Sandstede, and T.~Wagenknecht}, {\em
  Snakes, ladders, and isolas of localized patterns}, SIAM Journal on
  Mathematical Analysis, 41 (2009), pp.~936--972.

\bibitem{Belyakov1}
{\sc L.~Belyakov}, {\em A case of the generation of a periodic orbit motion
  with homoclinic curves}, Math. Notes, 15 (1980), pp.~336--341.

\bibitem{Belyakov2}
{\sc L.~Belyakov, L.~Glebsky, and L.~Lerman}, {\em Abundance of stable
  stationary localized solutions to the generalized 1d {Swift-Hohenberg}
  equation}, Comp. \& Math. App., 34 (1997), pp.~253--266.

\bibitem{Bordeu}
{\sc I.~Bordeu, M.~Clerc, R.~Lefever, and M.~Tlidi}, {\em From localized spots
  to the formation of invaginated labyrinthine structures in a
  {Swift-Hohenberg} model}, Commun. Nonlin. Sci. Num. Sim., 29 (2015),
  pp.~482--487.

\bibitem{Victor2}
{\sc V.~{Bre\~na--Medina} and A.~Champneys}, {\em Subcritical {Turing}
  bifurcation and the morphogenesis of localized patterns}, Phys. Rev. E, 90
  (2014), p.~032923.

\bibitem{Buffoni}
{\sc B.~Buffoni, A.~Champneys, and J.~Toland}, {\em Bifurcation and coalescence
  of a plethora of homoclinic orbits for a hamiltonian system}, Journal of
  Dynamics and Differential Equations, 8 (1996), pp.~221--279.

\bibitem{BurkeDawes}
{\sc J.~Burke and J.~Dawes}, {\em Localized states in an extended
  {Swift-Hohenberg} equation}, SIAM J. Appl. Dyn. Sys., 11 (2012),
  pp.~261--284.

\bibitem{Houghton1}
{\sc J.~Burke and E.~Houghton, S.M. amd~Knobloch}, {\em {Swift-Hohenberg}
  equation with broken reflection symmetry}, Phys Rev E, 80 (2009),
  p.~Art.~No:036202.

\bibitem{burke01}
{\sc J.~Burke and E.~Knobloch}, {\em Snakes and ladders: Localized states in
  the {S}wift--{H}ohenberg equation}, Physics Letters A, 360 (2006),
  pp.~681--688.

\bibitem{burke02}
{\sc J.~Burke and E.~Knobloch}, {\em Homoclinic snaking: Structure and
  stability}, Chaos, 17 (2007), p.~art.~no.~037102.

\bibitem{subsidiaryalan}
{\sc A.~Champneys}, {\em Subsidiary homoclinic orbits to a saddle-focus for
  reversible systems}, International Journal of Bifurcation and Chaos, 04
  (1994), pp.~1447--1482.

\bibitem{apd}
{\sc A.~Champneys}, {\em Homoclinic orbits in reversible systems and their
  applications in mechanics, fluids and optics}, Physica D: Nonlinear
  Phenomena, 112 (1998), pp.~158 -- 186.

\bibitem{apd2}
{\sc A.~Champneys}, {\em Homoclinic orbits in reversible systems {II}:
  {M}ulti-bumps and saddle-centres.}, CWI Quarterly, 12 (1999), pp.~185--212.

\bibitem{ChampWagen}
{\sc A.~Champneys, E.~Knobloch, Y.-P. Ma, and T.~Wagenknecht}, {\em Homoclinic
  snakes bounded by a saddle-center periodic orbit}, SIAM J. Appl. Dyn. Sys.,
  11 (2012), pp.~1583--1613.

\bibitem{Kozyreff2}
{\sc S.~Chapman and G.~Kozyreff}, {\em Exponential asymptotics of localised
  patterns and snaking bifurcation diagrams}, Physica D, 238 (2009),
  pp.~319--354.

\bibitem{Claudiohole}
{\sc M.~Clerc and C.~Falcon}, {\em Localized patterns and hole solutions in
  one-dimensional extended systems}, Physica A: Statistical Mechanics and its
  Applications, 356 (2005), pp.~48 -- 53.

\bibitem{Dawes2}
{\sc J.~Dawes}, {\em Localized pattern formation with a large-scale mode:
  Slanted snaking}, SIAM J. App. Dyn. Sys., 7 (2008), pp.~186--206.

\bibitem{Dawes}
{\sc J.~Dawes}, {\em Modulated and localised states in a finite domain}, SIAM
  J. Appl. Dyn. Syst., 8 (2009), pp.~909--930.

\bibitem{Dean}
{\sc A.~Dean, P.~Matthews, S.~Cox, and J.~King}, {\em Exponential asymptotics
  of homoclinic snaking}, Nonlinearity, 24 (2011), pp.~3323--3351.

\bibitem{Devaney1}
{\sc R.~Devaney}, {\em Homoclinic orbits in {H}amiltonian systems}, J. Diff.
  Eqns., 21 (1976), pp.~431--438.

\bibitem{Devaney2}
{\sc R.~Devaney}, {\em Blue sky catastrophes in reversible and {H}amiltonian
  systems}, Indiana Univ. Math. J., 26 (1977), pp.~247--263.

\bibitem{auto}
{\sc E.~Doedel, A.~Champneys, T.~Fairgrieve, Y.~Kuznetsov, B.~Sandstede, and
  X.~Wang}, {\em Auto 97: Continuation and bifurcation software for ordinary
  differential equations (with homcont)}, 2002.
\newblock Technical report, Concordia University.

\bibitem{doeloriginal}
{\sc A.~Doelman, T.~J. Kaper, and P.~A. Zegeling}, {\em Pattern formation in
  the one-dimensional gray - scott model}, Nonlinearity, 10 (1997), p.~523.

\bibitem{fritsexplicit}
{\sc A.~Doelman and F.~Veerman}, {\em An explicit theory for pulses in two
  component, singularly perturbed, reaction--diffusion equations}, Journal of
  Dynamics and Differential Equations, 27 (2015), pp.~555--595,
  \url{https://doi.org/10.1007/s10884-013-9325-2},
  \url{https://doi.org/10.1007/s10884-013-9325-2}.

\bibitem{ioosbook}
{\sc M.~Haragus and G.~Iooss}, {\em Local Bifurcations, Center Manifolds, and
  Normal Forms in Infinite-Dimensional Dynamical Systems}, Springer, 2011.

\bibitem{Harterich}
{\sc J.~Harterich}, {\em Cascades of homoclinic orbits to a saddle focus
  equilibrium}, Physica D, 112 (1998), pp.~187--200.

\bibitem{Houghton2}
{\sc S.~Houghton and E.~Knobloch}, {\em {Swift-Hohenberg} equation with broken
  cubic-quintic nonlinearity}, Phys Rev E, 84 (2011), p.~Art.~No:016204.

\bibitem{wardfirst}
{\sc D.~Iron, M.~J. Ward, and J.~Wei}, {\em The stability of spike solutions to
  the one-dimensional gierer–meinhardt model}, Physica D: Nonlinear
  Phenomena, 150 (2001), pp.~25 -- 62.

\bibitem{Knoblochreview}
{\sc E.~Knobloch}, {\em Spatial localization in dissipative systems}, Annual
  Review of Condensed Matter Physics, 6 (2015), pp.~325--359.

\bibitem{norevsnake}
{\sc J.~Knobloch, T.~Rie{\ss}, and M.~Vielitz}, {\em Nonreversible homoclinic
  snaking}, Dynamical Systems, 26 (2011), pp.~335--365.

\bibitem{Kozyreff}
{\sc G.~Kozyreff and S.~Chapman}, {\em Asymptotics of large bound state of
  localised structures}, Physical Review Letters, 97 (2006), p.~art.no.044502.

\bibitem{Kozyreff3}
{\sc G.~Kozyreff and S.~Chapman}, {\em Analytical results for front pinning
  between an hexagonal pattern and a uniform state in pattern-formation
  systems}, Phys. Rev. Lett., 111 (2013), p.~art.~no.~054501.

\bibitem{2Dsnake}
{\sc D.~Lloyd, B.~Sandstede, D.~Avitabile, and A.~Champneys}, {\em Localized
  hexagon patterns of the planar {Swift-Hohenberg} equation}, SIAM J. Appl.
  Dyn. Syst., 7 (2008), pp.~1049--1100.

\bibitem{crimelloyd}
{\sc D.~J. Lloyd and H.~O'Farrell}, {\em On localised hotspots of an urban
  crime model}, Physica D: Nonlinear Phenomena, 253 (2013), pp.~23 -- 39.

\bibitem{Defect}
{\sc Y.-P. Ma, J.~Burke, and E.~Knobloch}, {\em Defect-mediated snaking: A new
  growth mechanism for localized structures}, Physica D., 239 (2010),
  pp.~1867--1883.

\bibitem{Makrides}
{\sc E.~Makrides and B.~Sandstede}, {\em Predicting the bifurcation structure
  of localized snaking patterns}, Physica D, 26 (2014), pp.~59--78.

\bibitem{McCalla}
{\sc S.~McCalla and B.~Sandstede}, {\em Spots in the {Swift-Hohenberg}
  equation}, SIAM J. Appl. Dyn. Syst., 12 (2013), pp.~831--877.

\bibitem{Promislow}
{\sc R.~Moore and K.~Promislow}, {\em Renormalization group reduction of pulse
  dynamics in thermally loaded optical parametric oscillators}, Physica D, 206
  (2005), pp.~62--81.

\bibitem{Morior}
{\sc Y.~Mori, A.~Jilkine, and L.~{Edelstein-Keshet}}, {\em Wave-pinning and
  cell polarity from a bistable reaction-diffusion system}, Biophysical
  Journal, 94 (2008), pp.~3684 -- 3697.

\bibitem{Oldeman}
{\sc B.~Oldeman, B.~Krauskopf, and A.~Champneys}, {\em Numerical unfoldings of
  codimension-three resonant homoclinic flip bifurcations}, Nonlinearity, 14
  (2001), pp.~597--621.

\bibitem{parra2}
{\sc P.~Parra-Rivas, D.~Gomila, L.~Gelens, and E.~Knobloch}, {\em Bifurcation
  structure of localized states in the {Lugiato-Lefever} equation with
  anomalous dispersion}, Physical Review E, 97 (2018), p.~art.~no.~042204.

\bibitem{purwins}
{\sc H.~Purwins, H.~B{\"o}deker, and S.~Amiranashvili}, {\em Dissipative
  solitons}, Advances in Physics, 59 (2010), pp.~485--701.

\bibitem{wardrecent}
{\sc I.~Rozada, S.~J. Ruuth, and M.~J. Ward}, {\em The stability of localized
  spot patterns for the brusselator on the sphere}, SIAM Journal on Applied
  Dynamical Systems, 13 (2014), pp.~564--627.

\bibitem{Sevryuk}
{\sc M.~Sevryuk}, {\em Reversible Systems}, Springer, New York, 2009.
\newblock Lecture Notes in Mathematics.

\bibitem{shilnibook}
{\sc L.~P. Shil'nikov, L.~Shil'nokov~Andrey, D.~V. Turaev, and O.~Chua, Leon},
  {\em Methods of Qualitative Theory in Nonlinear Dynamics: Part I and II},
  World Scientific, 1998.

\bibitem{shortcrime}
{\sc M.~B. Short, M.~R. D'Orsonga, V.~B. Pasour, G.~E. Tita, P.~J. Brantingham,
  A.~L. Bertozzi, and L.~B. Chayes}, {\em A statistical model of criminal
  behavior}, Mathematical Models and Methods in Applied Sciences, 18 (2008),
  pp.~1249--1267.

\bibitem{miophil}
{\sc N.~Verschueren, U.~Bortolozzo, M.~Clerc, and S.~Residori}, {\em Chaoticon:
  localized pattern with permanent dynamics}, Philosophical Transactions of the
  Royal Society of London A: Mathematical, Physical and Engineering Sciences,
  372 (2014).

\bibitem{miosiads}
{\sc N.~Verschueren and A.~Champneys}, {\em A model for cell polarization
  without mass conservation}, SIAM Journal on Applied Dynamical Systems, 16
  (2017), pp.~1797--1830.

\bibitem{alansnake}
{\sc P.~Woods and A.~Champneys}, {\em Heteroclinic tangles and homoclinic
  snaking in the unfolding of a degenerate reversible {H}amiltonian--{H}opf
  bifurcation}, Physica D: Nonlinear Phenomena, 129 (1999), pp.~147--170.

\bibitem{yuvalss}
{\sc Y.~R. Zelnik, H.~Uecker, U.~Feudel, and E.~Meron}, {\em Desertification by
  front propagation?}, Journal of Theoretical Biology, 418 (2017), pp.~27 --
  35.

\end{thebibliography}

\end{document}